**Cavity Loss in QED Coupled-Cluster Theory: Polariton Linewidths, Driven Steady States, and the Rabi Splitting**

Kurt R. Brorsen, Department of Chemistry, University of Missouri, Columbia, Missouri 65211, United States (brorsenk@missouri.edu)

## Abstract

Molecular polariton experiments are performed in lossy cavities, whereas correlated *ab initio* polariton theory has been developed almost entirely for lossless cavities. In this work, cavity loss is added to real-time QED coupled-cluster theory (QED-CCSD-22) through a complex cavity frequency, which gives the non-Hermitian no-jump part of the Lindblad master equation. The resulting polariton linewidths and driven-dissipative steady states are, to our knowledge, the first from QED coupled-cluster theory. The linewidths obtained from the complex Jacobian and from real-time propagation agree and are partitioned by the biorthogonal photon numbers of the two polaritons. A closed-form continuous-wave drive of the cavity gives driven steady states and their lineshapes.

## 1. Introduction

Molecular polaritons form when a molecular transition couples strongly to a confined electromagnetic mode, and they have been used to modify chemical reactivity and photophysics.[1,2] Most experiments that create them are performed in lossy cavities. In plasmonic nanocavities, which reach single-molecule strong coupling at room temperature, the cavity decay rate $\kappa$ is comparable to or larger than the light–matter coupling $g$.[3–5] Measured quantities such as transmission spectra, polariton linewidths, and lifetimes therefore depend on the loss as much as on the coupling. Recent experiments have also compensated the loss in the analysis of measured spectra, by temporal deconvolution or complex-frequency excitation, to reveal polariton states and couplings that the linewidth hides.[6,7]

*Ab initio* polaritonic methods describe the molecule using electronic structure theory. In particular, QED Hartree–Fock and QED coupled-cluster (QED-CC) theory treat the correlated molecule and a quantized cavity mode on the same footing.[8–12] These methods have been reviewed[13] and implemented at scale,[14] QED-CC has been applied to molecules coupled to quantized plasmons,[15] and real-time QED-CC has been used to follow the electron–photon dynamics of molecules under laser pulses.[16] However, all of these studies compute the polaritons of a lossless cavity mode.

Many open-system treatments of molecular polaritons have used reduced descriptions of the molecule. Lindblad master equations with model or potential-energy-surface matter are standard,[17–20] and stochastic Schrödinger equations have been combined with molecules described by *ab initio* configuration interaction.[21] Non-Hermitian molecular dynamics has been applied to collective ensembles,[22] and Lindblad decay has been built into mean-field Ehrenfest dynamics of a molecule in a lossy cavity.[23]

Dissipative QED density-functional theory[24,25] and real-time nuclear–electronic orbital density-functional theory with lossy classical modes[26] have treated real molecules. Dissipative and driven coupled-cluster methods exist outside cavity QED, for electronic transport and thermal problems,[27–30] and equation-of-motion CC methods with complex absorbing potentials or complex scaling describe metastable electronic states through non-Hermitian Hamiltonians.[31,32] Closest to the present work, Antoniou et al. and McTague and Foley described cavity loss through a complex photon energy, the latter within configuration interaction singles.[33–35]

Recently, Bruno et al. included cavity loss in QED-CCSD by describing the cavity field as a weakly damped harmonic oscillator and averaging the resulting Hamiltonian over a time interval, which rescales the bilinear coupling and the dipole self-energy.[36] For p-nitroaniline, they found that the Rabi splitting, defined as the minimum gap between the two polariton eigenvalues of this Hermitian Hamiltonian, decreases at a cavity quality factor of 60. Because their effective Hamiltonian is Hermitian, it does not give polariton linewidths, and neither linewidths nor driven-dissipative steady states have been reported from QED-CC theory so far.

In this work, we add cavity loss to real-time QED-CCSD-22 through a complex cavity frequency, $\omega \to \omega - i\kappa/2$, in the photon energy, as in Refs. 33 and 34. The resulting non-Hermitian Hamiltonian generates the no-jump evolution of the Lindblad equation for a lossy cavity mode. We obtain polariton linewidths from the complex Jacobian and from real-time propagation, derive a closed-form continuous-wave drive of the cavity that gives driven-dissipative steady states, and determine which observables the no-jump description gives, and up to which drive strength, by comparison with exact Lindblad dynamics of $H_2$. The same exact reference is then used to compare the quantities that have been called a dissipative Rabi splitting.

## 2. Theory

### 2.1 Real-time QED-CCSD-22

For a molecule coupled to a single cavity mode, the Pauli–Fierz Hamiltonian in the length gauge and the coherent-state basis is[8]

$$\hat{H} \;=\; \hat{H}_{\mathrm{el}} \;+\; \omega\, \hat{b}^\dagger \hat{b} \;-\; \sqrt{\omega/2}\, (\boldsymbol{\lambda} \cdot \hat{\boldsymbol{d}})\, (\hat{b}^\dagger + \hat{b}) \;+\; 1/2\, (\boldsymbol{\lambda} \cdot \hat{\boldsymbol{d}})^2, \tag{1}$$

where $\hat{H}_{\mathrm{el}}$ is the electronic Hamiltonian, $\omega$ is the cavity frequency, $\boldsymbol{\lambda}$ is the coupling vector, and $\hat{b}^\dagger$ and $\hat{b}$ are the photon creation and annihilation operators, respectively. The dipole operator is shifted by its QED Hartree–Fock (QED-HF) expectation value, $\hat{\boldsymbol{d}} = \hat{\boldsymbol{\mu}} - \langle \hat{\boldsymbol{\mu}} \rangle$, and the last term is the dipole self-energy. In this study, the correlated wavefunction is QED-CCSD-22,[11,12] whose cluster operator contains electronic single and double excitations, the photon excitations $\hat{b}^\dagger$ and $\hat{b}^{\dagger 2}$, and all products of electronic excitations with up to two photons.

The cluster amplitudes $t_\mu$ are propagated with the time-dependent coupled-cluster equations,[16]

$$i\, \frac{dt_\mu}{dt} \;=\; \Omega_\mu(\boldsymbol{t}) \;=\; \langle \mu |\, e^{-\hat{T}} \hat{H}\, e^{\hat{T}} | \mathrm{HF} \rangle, \tag{2}$$

where $\langle \mu |$ runs over the projection manifold, with $\langle \mu | \hat{\tau}_\nu | \mathrm{HF} \rangle = \delta_{\mu\nu}$ for the excitation operators $\hat{\tau}_\nu$ of $\hat{T}$. Real-time coupled-cluster methods of this kind are well established for cavity-free molecules.[37–41] We propagate the deviation from the stationary state with a fourth-order exponential time-differencing integrator (ETDRK4).[42,43] The working equations were derived symbolically with SASQ[44] and are evaluated in a locally developed complex-arithmetic code built on PySCF.[45] The Supporting Information describes how the generated residuals of the two-photon amplitudes are normalized for the propagation.

### 2.2 Cavity loss through a complex cavity frequency

For a cavity mode that loses photons at the rate $\kappa$, the Lindblad master equation is

$$\frac{d\hat{\rho}}{dt} \;=\; -i[\hat{H}, \hat{\rho}] + \kappa\big(\hat{L}\hat{\rho}\hat{L}^\dagger - 1/2\, \{\hat{L}^\dagger \hat{L}, \hat{\rho}\}\big), \tag{3}$$

where $\hat{L}$ is the collapse operator. The term $\kappa \hat{L}\hat{\rho}\hat{L}^\dagger$ is the jump term, which returns population to lower states. The remaining terms combine with $\hat{H}$ into the non-Hermitian effective Hamiltonian

$$\hat{H}_{\text{eff}} = \hat{H} - i\frac{\kappa}{2}\hat{L}^\dagger\hat{L}, \tag{4}$$

which generates the no-jump evolution.[46–48] For $\hat{L} = \hat{b}$, Eq. (4) follows from Eq. (1) by the substitution $\omega \to \omega - i\kappa/2$ in the photon energy. The cavity frequency enters the QED-CCSD-22 residuals only as the coefficient of the photon-diagonal terms, so making it complex gives the lossy theory without any other change. The bilinear coupling is kept at the real frequency, unlike in Ref. 34, because a complex prefactor $\sqrt{\omega/2}$ would add a non-Hermitian term that is not of the form $\hat{b}^\dagger\hat{b}$. In QED-CIS, this term gives the ground state a positive imaginary energy, so that its norm grows, and changes the LP and UP linewidths of $H_2$ by $-6\%$ and $+7\%$ (Supporting Information).

In the coherent-state basis of Eq. (1), the mode that decays is the displaced mode, which in terms of the original photon operator is $\hat{b} - \langle\hat{b}\rangle$ with the QED-HF displacement $\langle\hat{b}\rangle$, so the mean-field ground state does not radiate. With the electronic dipole operator used here, $\langle\hat{b}\rangle$ depends on the origin of the coordinates, whereas the displaced mode does not. With the full dipole operator, $\langle\hat{b}\rangle$ depends on the origin only for charged molecules.[8] For both molecules studied here, the QED-HF displacement equals the exact $\langle\hat{b}\rangle$ by symmetry. For a polar molecule with $\boldsymbol{\lambda}$ along its dipole moment, the two differ by the correlation correction to the dipole moment, and the damped mode differs from $\hat{b} - \langle\hat{b}\rangle$ by a constant. For $H_2O$ with $\boldsymbol{\lambda}$ along its symmetry axis, this difference makes the imaginary part of the stationary energy about 1% more negative and broadens the polariton linewidths by less than $5 \times 10^{-4}$ relative (Supporting Information). We report photon populations as expectation values of $\hat{n}' = (\hat{b} - \langle\hat{b}\rangle)^\dagger(\hat{b} - \langle\hat{b}\rangle)$, the photon number of the displaced mode, which does not depend on the origin.

The correlated ground state of $\hat{H}$ is not stationary under $\hat{H}_{\text{eff}}$, so we converge the stationary state of the lossy equations with a quasi-Newton solver with DIIS acceleration, started from the Hermitian ground state, and all kicked and driven trajectories start from this complex stationary state. Linear response about the stationary state is described by the complex Jacobian $\partial\Omega_\mu/\partial t_\nu$, whose eigenvalues $\omega_k - i\gamma_k$ give the polariton frequencies and linewidths, with $\gamma_k$ the half width at half maximum.

### 2.3 Continuous-wave drive of the cavity

A continuous-wave drive of the cavity mode adds $\eta(t)\,(\hat{b}^\dagger + \hat{b})$ to $\hat{H}$, where $\eta(t)$ is the drive amplitude. Because $\hat{b}^\dagger$ commutes with $\hat{T}$ and the commutator $[\hat{b}, \hat{T}]$ contains only excitation operators, which also commute with $\hat{T}$, the similarity transformation of the drive terminates after one commutator. With the normalization of the amplitudes used here, the drive adds the terms

$$\begin{gathered} \Omega_{t_1} \mathrel{+}= \eta\, s_1^1, \quad \Omega_{t_2} \mathrel{+}= \eta\, s_2^1, \quad \Omega_{p_1} \mathrel{+}= \eta\,(1 + p_2), \\ \Omega_{s_1^1} \mathrel{+}= \eta\, s_1^2, \quad \Omega_{s_2^1} \mathrel{+}= \eta\, s_2^2, \quad \text{two-photon rows } 0, \end{gathered} \tag{5}$$

to the residuals, where $t_1$ and $t_2$ are the electronic amplitudes, $p_1$ and $p_2$ the one- and two-photon amplitudes, and $s_n^m$ the amplitudes that combine $n$ electronic excitations with $m$ photons. The integrals do not change during the propagation. An independent construction that absorbs the drive into the one-electron integrals gives the same terms (Supporting Information).

### 2.4 Photon observables

Photon observables are computed from the raw moments $M_{\hat{O}} = \langle\text{HF}|(1 + \hat{\Lambda})\, e^{-\hat{T}}\hat{O}\, e^{\hat{T}}|\text{HF}\rangle$ of Ref. 49, evaluated with the time-dependent $\hat{T}(t)$, which starts from the complex stationary state,

and with $\widehat{\Lambda}$ fixed at the ground-state value $\widehat{\Lambda}_0$ of the lossless cavity. In the ground state, such moments give the intracavity photon statistics and the vacuum-field squeezing of cavity-coupled molecules.[49,50] These moments carry the same poles and linewidths as the dipole signal. However, they contain no contribution that is quadratic in the field, because every term of $M_{\hat{b}^\dagger\hat{b}}$ contains a photon $\Lambda$ amplitude, so the photon occupation computed this way vanishes at $\lambda = 0$. We therefore obtain the coherent part of the photon number as

$$n_{\mathrm{coh}}(t) \;=\; |\delta M_{\hat{b}}(t)|^2, \qquad \delta M_{\hat{b}} = M_{\hat{b}} - M_{\hat{b}}^{\mathrm{GS}}, \tag{6}$$

which equals $\langle\hat{n}\rangle(t)$ at $\lambda = 0$ because no-jump loss maps a coherent state onto a coherent state. The coupled-cluster estimate of $\langle\hat{n}'\rangle(t)$ is the coupled-cluster ground-state value, the incoherent photon number of Ref. 49, plus $n_{\mathrm{coh}}(t)$. Photon-number populations beyond the coherent part require a time-dependent $\widehat{\Lambda}$ and are not computed in this work.

The left factor $\langle\mathrm{HF}|(1+\widehat{\Lambda}_0)\, e^{-\hat{T}(t)}$ of these moments is neither the exact ground state nor the adjoint of the propagated state, so the moments are not normalized expectation values. In Sec. 4.5, the coupled-cluster coherence $\delta M_{\hat{b}}$ and photon-number estimate are therefore compared with two exact quantities for $H_2$, the normalized expectation value $\langle\psi|\hat{O}|\psi\rangle/\langle\psi|\psi\rangle$ in the no-jump state $|\psi(t)\rangle$ and the expectation value $\mathrm{Tr}[\hat{O}\hat{\rho}(t)]$ of the Lindblad dynamics. Table 1 compares the two exact quantities with each other, and Fig. 5 compares both with the coupled-cluster results.

## 3. Computational details

We consider $H_2$/6-31G ($R = 0.74$ Å) with $\lambda = 0.05$ a.u. along the bond and $\omega = 0.562595$ Ha, and $H_2O$/6-31G (O at the origin, H at $(0, \pm 0.757, 0.587)$ Å) with $\lambda = 0.05$ a.u. along the out-of-plane axis and $\omega = 0.308197$ Ha. In both cases the cavity is set in resonance with the lowest excitation that couples to the cavity polarization at $\lambda = 0$, which is the $1b_1 \to 4a_1$ excitation for $H_2O$. At $\lambda = 0.05$ a.u. the dipole self-energy shifts this excitation, which leaves the cavity below the dressed excitation by $0.21g$ for $H_2$ and $0.33g$ for $H_2O$, where $g$ is the coupling of the two-level model of Sec. 4.3. The loss rates $\kappa = 0.002$–$0.03$ a.u. ($0.05$–$0.82$ eV) correspond to quality factors $Q = \omega/\kappa$ down to 28 for the $H_2$ linewidths and 10 for $H_2O$, as in plasmonic nanocavities.[3–5] For the comparison of Sec. 4.3, the $H_2$ range is extended to $Q = 5$. We note that molecules in plasmonic gaps couple to the field through the Coulomb interaction, without a dipole self-energy,[51] so the present calculations use plasmonic loss rates but do not model a plasmonic cavity.

For p-nitroaniline, we use the geometry, the basis set, the coupling strength, and the cavity frequency of Ref. 36: cc-pVDZ, $\lambda = 0.05$ a.u. along the long molecular axis, and $\omega = 5.660$ eV, at which their lossless gap is smallest, with all electrons correlated. Only states of $A_1$ symmetry couple to the cavity, so the complex polariton roots are obtained by a Davidson solver in this symmetry sector, with products of the Jacobian and a vector computed by central differences of the residual (Supporting Information).

Kicked trajectories use a time step of 0.5 a.u. and run to 800 a.u. ($H_2$) or 1600 a.u. ($H_2O$), and driven trajectories run to 500 a.u. at each drive frequency. Each time step requires four evaluations of the QED-CCSD-22 residual in complex arithmetic, which scale as the ground-state equations ($N^6$). Linewidths are extracted from the real-time signals by a variable-projection fit to damped cosines, seeded by the peaks of a damped Fourier transform (Supporting Information). For $H_2O$, the complex stationary state and the complex Jacobian are computed at twenty loss rates between $\kappa = 0.002$ and 0.03 a.u., and the two polariton roots are followed by continuation in $\kappa$.

The exact reference is the Pauli–Fierz Hamiltonian of $H_2$/6-31G in the product of the full configuration-interaction electronic space and a photon Fock space truncated at $n_{\max} = 8$ (144 states). Raising $n_{\max}$ to 10 changes the polariton roots by less than $10^{-12}$ Ha. Because $H_2$ has two electrons, coupled-cluster theory is exact in the electronic space, and deviations from this reference measure the truncation of the photon excitation rank. The Lindblad equation is integrated by fourth-order Runge–Kutta on the density matrix, and the no-jump evolution is obtained by diagonalizing $\hat{H}_{\text{eff}}$.

Two collapse operators are compared. The first is the displaced mode $\hat{L} = \hat{b} - \langle \hat{b} \rangle$, which for $H_2$ is the operator that the coupled-cluster propagation damps. The second is a dressed operator that connects the eigenstates of $\hat{H}$ with the rates of an ohmic bath spectral density,[52–55] for which the correlated ground state is exactly dark, that is, annihilated by the collapse operator. This operator is the jump operator of the universal Lindblad equation of Nathan and Rudner[56,57] for a bath at zero temperature, without the Lamb shift, and because it is a single operator, the dissipator keeps the cross terms between the decay channels of different transitions (Supporting Information). Both Lindblad models are solved numerically exactly for this Hamiltonian and these dissipators.

## 4. Results and discussion

### 4.1 Validation

Each part of the implementation was checked against an exact limit (Supporting Information). At $\lambda = 0$, the Jacobian reproduces the EOM-CCSD excitation energies, and its one- and two-photon roots lie at $\omega - i\kappa/2$ and $2\omega - i\kappa$. At $\lambda = 0.05$ a.u., the complex QED-CCSD-22 polariton roots of $H_2$ agree with the exact complex excitation energies $E_k - E_0$ of $\hat{H}_{\text{eff}}$, where $E_0$ belongs to the stationary state, to $1.4 \times 10^{-7}$ Ha for $\kappa$ up to 0.11 a.u. This is the accuracy of the lossless roots, so the remaining error comes from the truncation of the photon excitation rank.

The linewidths can be extracted from real-time propagation. For $H_2$ at $\kappa = 0.02$ a.u., damped-cosine fits of the kicked dipole signal and of the coherent photon displacement $\delta M_{\hat{b}}(t)$ recover the Jacobian linewidths of the lower and upper polaritons (LP and UP) to 0.1%. For $H_2O$ at $\kappa = 0.01$ a.u., where the polariton doublet is weak compared with nearby electronic peaks of the dipole spectrum, the fitted linewidths agree with the Jacobian to within a few percent once these electronic modes are included in the fit (Supporting Information).

### 4.2 Polariton linewidths

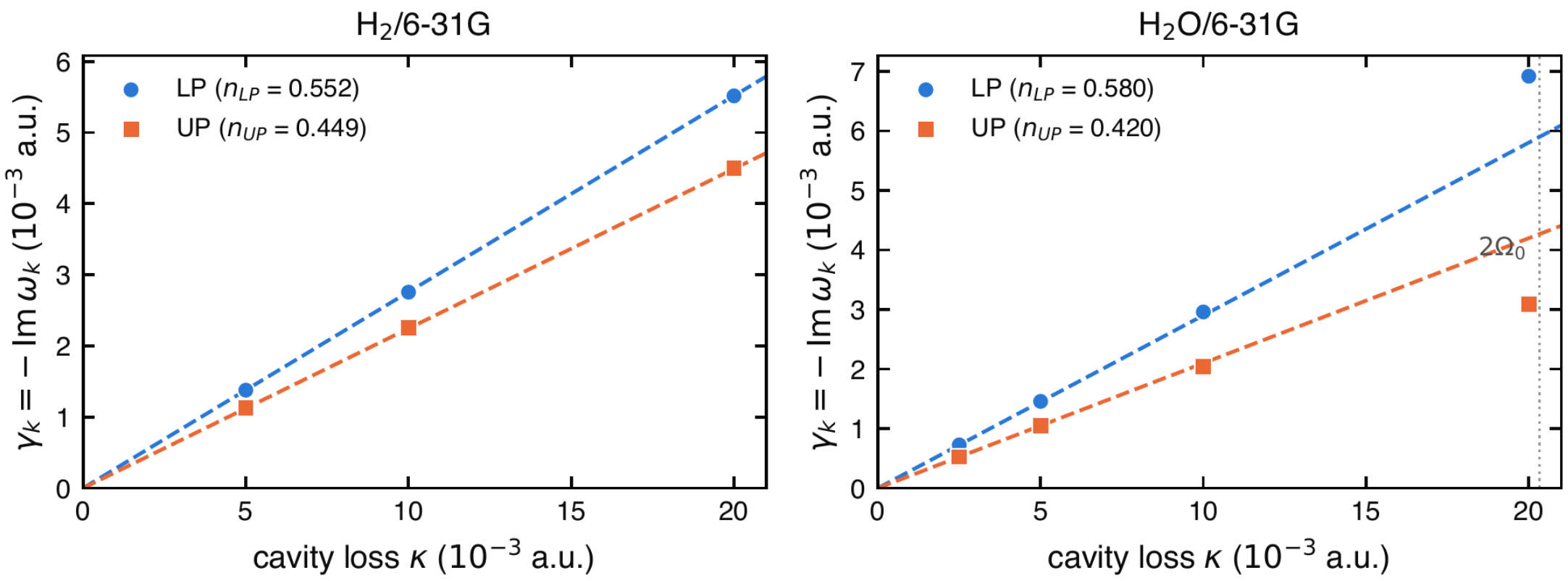

Figure 1: Polariton linewidths $\gamma_k = -\mathrm{Im}\,\omega_k$ from the complex Jacobian versus the cavity loss rate $\kappa$ for $H_2$/6-31G (left) and $H_2O$/6-31G (right) at $\lambda = 0.05$ a.u. Dashed lines show the first-order partition $\gamma_k = (\kappa/2)\,n_k$ with the biorthogonal photon number $n_k$ of the $\kappa = 0$ Jacobian. The dotted vertical line in the right panel marks $\kappa = 2\Omega_0$, the exceptional point of the resonant two-level model (Sec. 4.3).

Figure 1 shows the polariton linewidths of both molecules as a function of the loss rate. For $H_2$, the linewidths closely follow $\gamma_k = (\kappa/2)\,n_k$, where $n_k = (l_k|\hat{n}|r_k)/(l_k|r_k)$ is the biorthogonal photon number of the left and right eigenvectors of the $\kappa = 0$ Jacobian. This partition is first-order perturbation theory for the non-Hermitian Jacobian.[58] The small remaining deviation comes from the response of the stationary state to the loss, which the complex Jacobian includes. With this response, $\gamma_k/(\kappa/2)$ approaches the excess photon number $\langle k|\hat{n}|k\rangle - \langle \mathrm{G}|\hat{n}|\mathrm{G}\rangle$ of the exact polariton eigenstate $|k\rangle$ as $\kappa \to 0$ (Supporting Information). In contrast, the squared photon weight of the normalized right eigenvector, which is often quoted as the photon fraction, overestimates $\gamma_{\mathrm{LP}}$ by 28%, because the right eigenvectors of the nonsymmetric coupled-cluster Jacobian are not orthonormal.

For $H_2O$, whose Rabi splitting is about seven times smaller than that of $H_2$, the biorthogonal photon numbers partition the linewidths equally well at small loss rates, and the partition degrades as $\kappa$ grows. At $\kappa = 0.02$ a.u., a loss rate that leaves the $H_2$ doublet in the linear regime, the $H_2O$ doublet is close to the boundary between strong and weak coupling, which we discuss in the next section.

A lower-level treatment changes these linewidths through the polariton energies and photon numbers that it predicts, not through the second photon. QED-CCSD-21, which keeps one photon, reproduces the $H_2O$ linewidths and splittings of QED-CCSD-22 almost exactly. The non-Hermitian QED-CIS of Ref. 34, with the cavity in resonance with the excitation of each method, predicts Rabi splittings 5–10% larger than QED-CCSD-22, which is electronically exact for $H_2$, and linewidths that differ by a few percent (Supporting Information).

Finally, the linewidths depend on the model of the loss. A dressed collapse operator with an ohmic bath spectral density gives nearly the same linewidths as the bare-mode operator, whereas a flat spectral density changes them by about 6% (Supporting Information). The ohmic agreement follows from the commutator $[\hat{H}, \hat{X}] = \omega(\hat{b}^\dagger - \hat{b})$ of $\hat{X} = \hat{b}^\dagger + \hat{b}$ with Eq. (1), which gives

$$|\langle \mathrm{G}|\hat{b}|k\rangle|^2 - |\langle \mathrm{G}|\hat{b}^\dagger|k\rangle|^2 \;=\; \frac{\omega_k}{\omega}\,|\langle \mathrm{G}|\hat{X}|k\rangle|^2 \tag{7}$$

for an eigenstate $|k\rangle$ with excitation energy $\omega_k$. The right-hand side is the ohmic decay rate of $|k\rangle$ to the ground state divided by $\kappa$, and the left-hand side is the leading part of the excess photon number of $|k\rangle$. A Markovian single-mode model also excludes interference between several lossy modes, which can produce polaritons below the single-mode strong-coupling criterion.[59] At the low quality factors reached here, $Q \approx 15$ for $H_2O$ and down to 5 in Sec. 4.3, such a model should be compared with quantized quasinormal-mode theories, which describe broadband lossy cavities from first principles and set the limits of phenomenological dissipative models.[60,61]

### 4.3 Cavity loss and the Rabi splitting

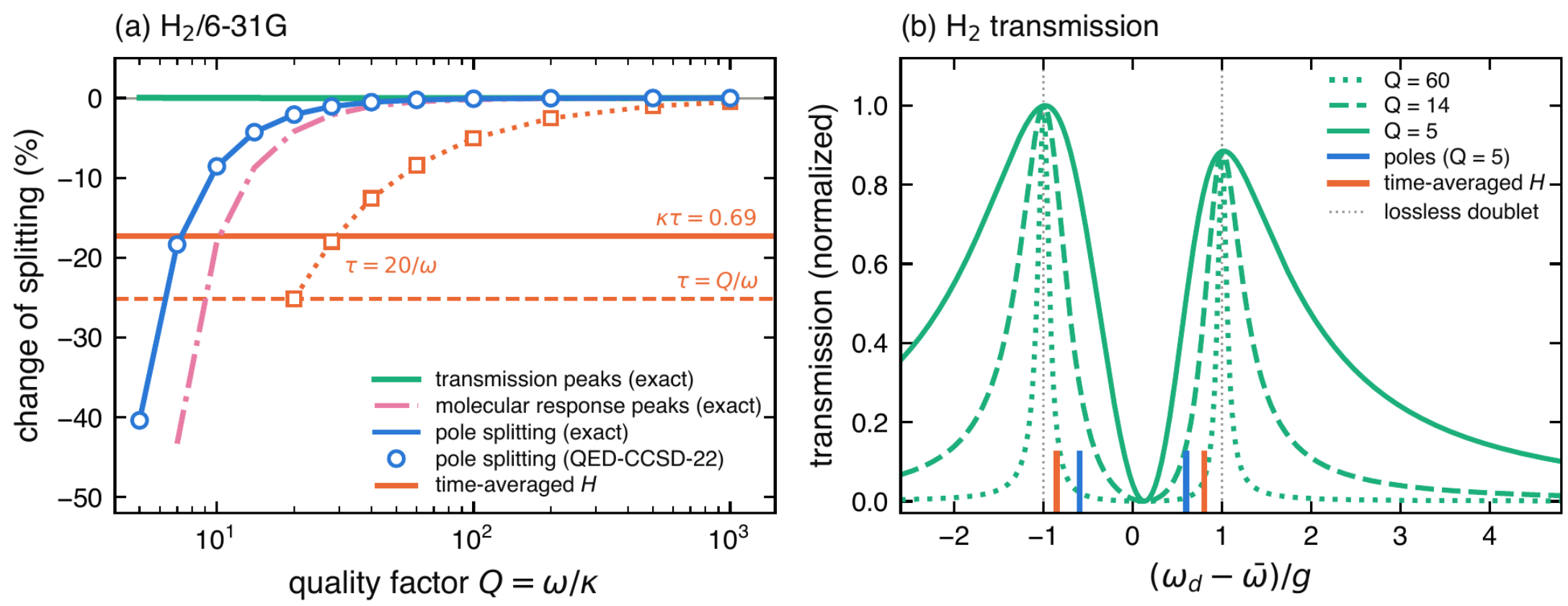

Figure 2: Effect of cavity loss on the Rabi splitting of $H_2$/6-31G ($\lambda = 0.05$ a.u.). (a) Relative change of the splitting versus the quality factor $Q = \omega/\kappa$ for the real part of the pole splitting of $\hat{H}_{\text{eff}}$ (exact with the dressed collapse operator, solid; QED-CCSD-22, circles), peak separations of the exact transmission spectrum and of the exact response of the molecular dipole, whose two peaks merge between $Q = 7$ and 5, and the eigenvalue gap of the time-averaged Hamiltonian of Ref. 36 for three choices of the averaging time $\tau$ (QED-CCSD-22 and exact values coincide). (b) Exact transmission spectra for $Q = 60$, 14, and 5, normalized to their maxima, with the pole positions at $Q = 5$ and the eigenvalues of the time-averaged Hamiltonian at $\kappa\tau = 0.69$ marked. The dotted lines mark the lossless polaritons. The frequency axis is centered on the mean of the lossless polaritons and scaled by $g = \Omega_0/2$.

Three quantities have been used as the Rabi splitting of a lossy cavity: the separation of the real parts of the two complex poles of $\hat{H}_{\text{eff}}$, the separation of the two peaks of a measured spectrum, and the gap between the polariton eigenvalues of a Hermitian Hamiltonian with loss-modified couplings. For a resonant two-level emitter with cavity loss only, the poles are $\bar{\omega} - i\kappa/4 \pm \sqrt{g^2 - \kappa^2/16}$, so the pole splitting changes only at second order in $\kappa/g$.[33] The peak positions of a spectrum depend on the detection channel,[62,63] because the driven and the undriven oscillator respond differently.[64] The maxima of the cavity transmission lie at $\bar{\omega} \pm g$ for every $\kappa$, which follows from the linear absorption and dispersion of the emitter inside the cavity,[65] whereas those of the emitter response to the same drive lie at $\bar{\omega} \pm (g^2 - \kappa^2/8)^{1/2}$. The time-averaged Hamiltonian of Ref. 36 multiplies the bilinear coupling by $1 - \kappa\tau/4$ and the dipole self-energy by $1 - \kappa\tau/2$, where $\tau$ is the averaging time, so its gap changes at first order in $\kappa\tau$.

Figure 2a compares these quantities for $H_2$ from $Q = 5$ to 1000, and the exact results confirm the two-level picture. The pole splitting decreases by 0.23% at $Q = 60$ and appreciably only at low $Q$. QED-CCSD-22 reproduces the exact values at every $Q$. The decrease comes from the cross term between the decay of the two polaritons into the same ground state, which the single collapse operator keeps. A secular master equation with the same rates, which treats the two decay channels as independent, leaves the pole splitting unchanged (Supporting Information).

The peaks of the exact transmission spectrum stay at their lossless positions down to $Q = 5$, where the pole splitting has decreased by 40%. Loss broadens these peaks without moving them, and because the transmission of a lossless emitter vanishes at the emitter frequency, the doublet does not merge (Fig. 2b). For an emitter with a linewidth of its own, the transmission maxima move slowly apart as $\kappa$ grows (Supporting Information). The response of the molecular dipole to the same drive, which a weak radiative channel of the molecule itself would detect, follows the emitter response instead. Its peak separation decreases at twice the rate of the pole splitting, and its two peaks merge at the lowest quality factors (Fig. 2a). The Lindblad equation used here omits the Lamb-shift term of the master equation, which for a frequency-dependent bath, including an ohmic one, shifts the two polaritons by different amounts at first order in $\kappa$.[48] Such shifts are usually absorbed into the measured cavity and transition frequencies.

In contrast, the gap of the time-averaged Hamiltonian is set by the averaging time, which Ref. 36 interprets as the time scale of the process being probed, and it depends on $Q$ only through $\kappa\tau = \omega\tau/Q$. Ref. 36 averaged over 5.01 fs at $Q = 60$, which gives $\kappa\tau = 0.69$ at 5.43 eV, near the resonance of p-nitroaniline with this Hamiltonian, and with this value the gap decreases by 17% for both $H_2$ and $H_2O$. With $\tau = Q/\omega$, the upper end of their averaging interval, the gap decreases by 25%. Taking the minimum gap over a scan of the cavity frequency, as Ref. 36 does, changes these values negligibly (Supporting Information).

For p-nitroaniline, we evaluated the time-averaged Hamiltonian with QED-CCSD-21, the method of Ref. 36, and the settings of its archived input files.[66] The scaled dipole self-energy moves the resonance to lower cavity frequency, and near it the gap decreases by 11.1%, less than for $H_2$ and $H_2O$ at the same $\kappa\tau$ (Supporting Information).

As for $H_2$, the pole splitting of p-nitroaniline changes little with loss. At the cavity frequency of Ref. 36 and $Q = 60$, QED-CCSD-22 gives a decrease of 0.05%, and the polaritons acquire full widths at half maximum of 51 and 42 meV, which add up to nearly the cavity linewidth. Even at $Q = 5$, the pole splitting decreases by only 7% (Supporting Information).

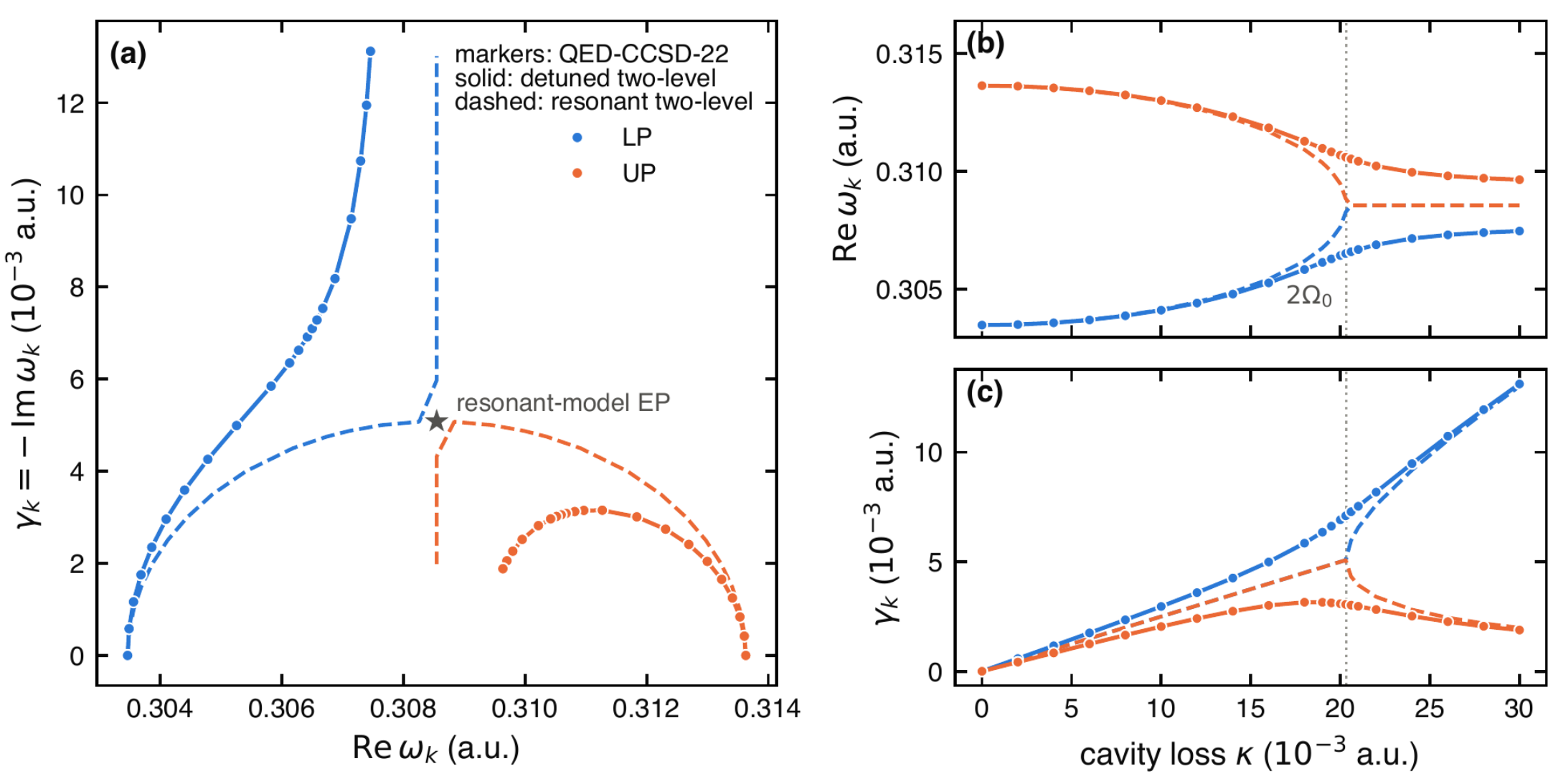

Figure 3: Polariton doublet of $H_2O$/6-31G ($\lambda = 0.05$ a.u.) as the loss rate approaches $2\Omega_0$. (a) Complex-Jacobian eigenvalues of the LP and UP in the complex frequency plane for $\kappa = 0.002$–$0.03$ a.u. (markers), together with the resonant two-level model (dashed), whose branches coalesce at the exceptional point (star), and the two-level model with the detuning fixed by the $\kappa = 0$ photon weights (solid). (b) Real parts and (c) linewidths $\gamma_k = -\mathrm{Im}\,\omega_k$ versus $\kappa$. The dotted line marks $\kappa = 2\Omega_0 = 0.0203$ a.u., the exceptional point of the resonant two-level model.

The pole splitting decreases strongly as $\kappa$ approaches the boundary between strong and weak coupling. For a resonant two-level system, this boundary is the exceptional point at which both the eigenvalues and the eigenvectors coalesce,[67,68] and such degeneracies have been realized across photonics and observed directly in exciton–polariton systems.[69–71] The $H_2O$ doublet is narrow, so $\kappa = 0.02$ a.u. ($Q \approx 15$) lies just below the resonant two-level value $\kappa_{\mathrm{EP}} = 2\Omega_0$. Figure 3 follows both complex roots through this region. At $\kappa = 0.02$ a.u., the real splitting has fallen below the linewidth of the LP, so that the two poles overlap, and the first-order partition breaks down. The two roots come closest there but do not coalesce at any real loss rate.

A two-level model explains this behavior once the detuning of Sec. 3 is included. With the detuning $\Delta = (n_{\mathrm{LP}} - n_{\mathrm{UP}})\,\Omega_0$ and the coupling $\sqrt{\Omega_0^2 - \Delta^2}/2$, both taken from the $\kappa = 0$ Jacobian, the model reproduces all twenty complex roots to about 0.2% (Fig. 3), whereas the resonant model underestimates the splitting at $\kappa = 0.02$ a.u. by more than a factor of two. The exceptional point of the detuned model lies at a complex loss rate, which no real loss rate reaches (Supporting Information).

### 4.4 Driven steady states

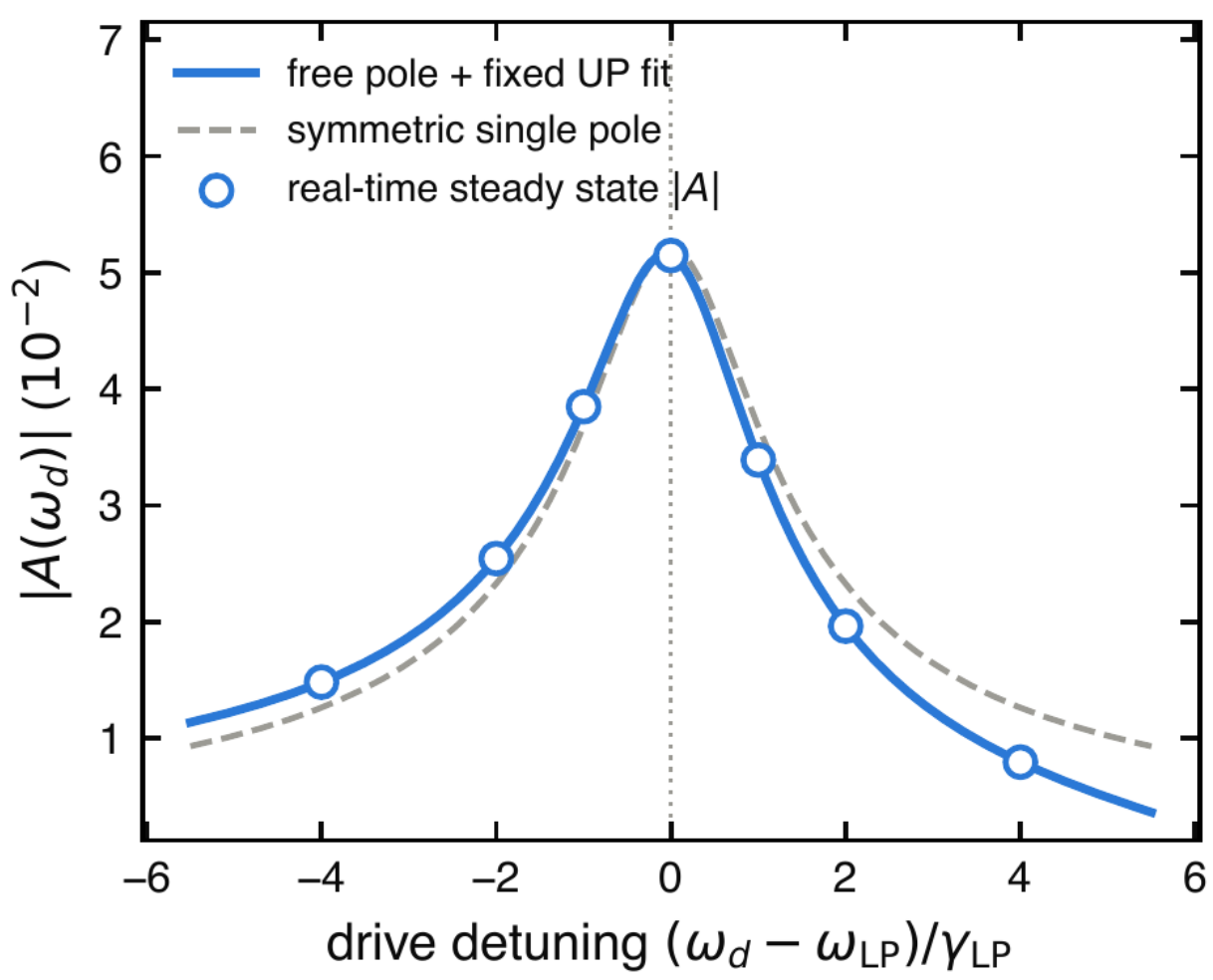

Figure 4: Driven steady-state lineshape across the lower polariton of $H_2$/6-31G ($\lambda = 0.05$ a.u., $\kappa = 0.02$ a.u., drive amplitude $\eta = 10^{-3}$). Markers show the co-rotating steady amplitude $|A(\omega_d)|$ of the one-photon amplitude $p_1$ from real-time propagation. The solid line is a fit of the complex amplitude with a free pole and the UP pole fixed at its Jacobian value, and the dashed line is a symmetric single-pole magnitude, which the data do not follow because the LP and UP responses interfere.

With the drive of Eq. (5), propagation reaches driven-dissipative steady states. A pumped cavity has also been described at the mean-field level by constraining the photon number of a strong-coupling QED-HF wave function.[72] We drove $H_2$ at $\kappa = 0.02$ a.u. with $\eta = 10^{-3}$ at seven frequencies around the LP and obtained the co-rotating steady amplitude $A(\omega_d)$ of the one-photon amplitude $p_1$ by a linear least-squares fit that includes the switch-on transient at the known Jacobian poles (Supporting Information). As shown in Fig. 4, $|A(\omega_d)|$ is not a symmetric Lorentzian, because the LP and UP responses interfere constructively below the LP and destructively between the polaritons. A fit of the complex amplitude to a free pole plus the UP pole fixed at its Jacobian value recovers the position and width of the LP pole closely (Supporting Information). Because the UP pole and the transient poles are taken from the Jacobian, this fit is a check of the steady amplitudes rather than an independent measurement of the linewidth. On resonance, the steady amplitude is 57% of the isolated-pole value $(\eta/2)/\gamma_{\mathrm{LP}}$, which is the residue $\langle \mathrm{G}|\hat{b}|\mathrm{LP}\rangle\langle \mathrm{LP}|\hat{X}|\mathrm{G}\rangle$ of the LP pole in the response of $\hat{b}$ to the drive.

### 4.5 Comparison with Lindblad dynamics

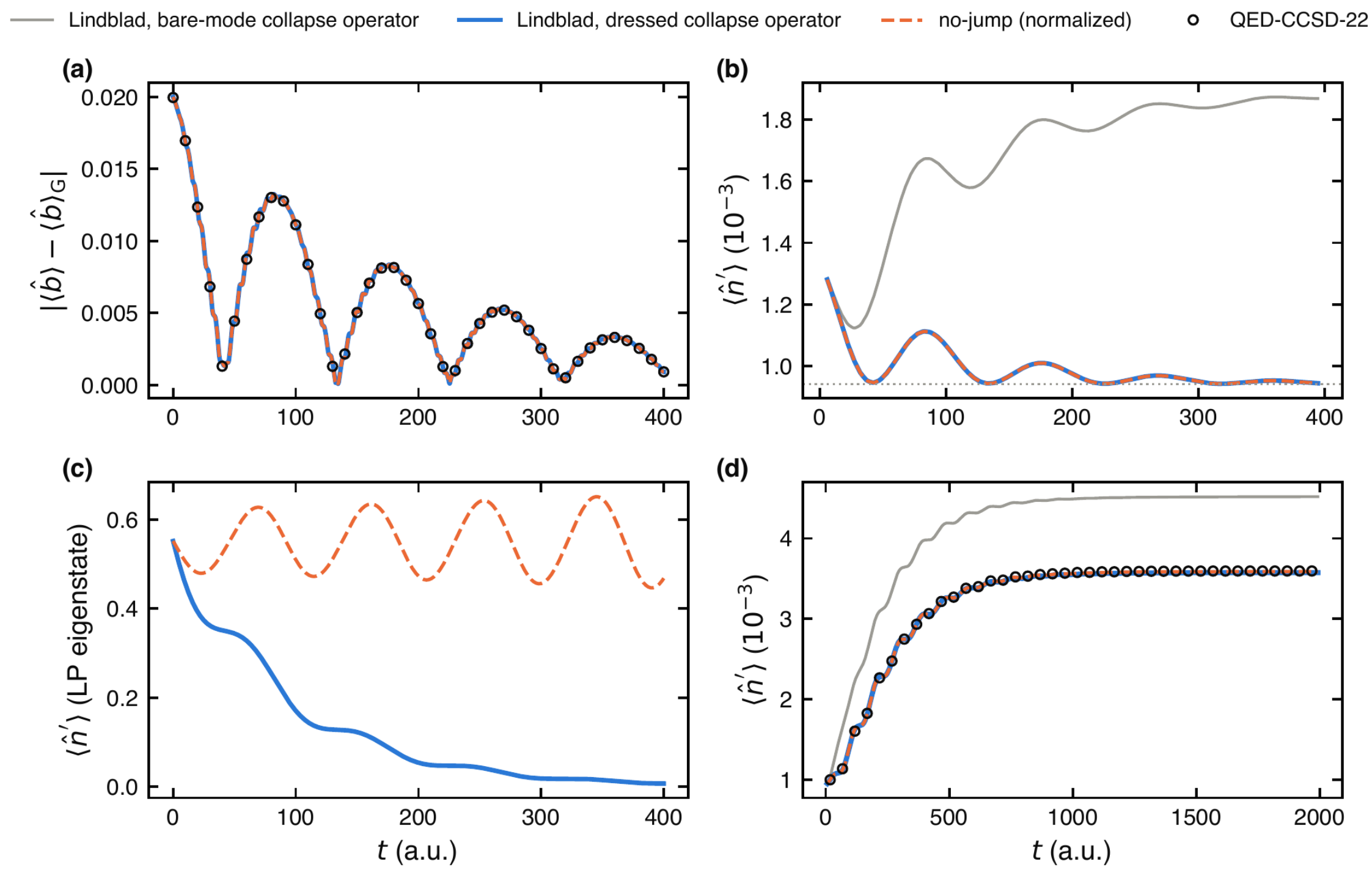

Figure 5: Exact Lindblad and no-jump dynamics of $H_2$/6-31G ($\lambda = 0.05$ a.u., $\kappa = 0.02$ a.u.). Dashed lines show the normalized no-jump evolution with the bare-mode collapse operator, solid lines the Lindblad dynamics with the dressed collapse operator, thin gray lines the Lindblad dynamics with the bare-mode collapse operator, and circles QED-CCSD-22. (a) Coherence $|\langle\hat{b}\rangle - \langle\hat{b}\rangle_{\mathrm{G}}|$ and (b) photon number of the displaced mode $\langle\hat{n}'\rangle$, averaged over one LP period, after a coherent kick $\varepsilon = 0.02$. The dotted line marks the ground-state value of $\langle\hat{n}'\rangle$, above which the bare-mode population relaxes because that dissipator does not leave the correlated ground state dark. (c) $\langle\hat{n}'\rangle$ after excitation into the lower-polariton eigenstate. (d) $\langle\hat{n}'\rangle$ averaged over one drive period under the drive of Fig. 4. For QED-CCSD-22, it is the ground-state value plus the coherent photon number $n_{\mathrm{coh}}$.

The complex frequency describes only the no-jump part of Eq. (3). Earlier comparisons found that a non-Hermitian Hamiltonian reproduces the linear spectra and coherences of quantum dots coupled to a plasmon.[73] For a polyatomic molecule in a laser-pumped lossy cavity, the same description was found to reproduce the polariton dynamics provided the pump is moderate and not much longer than the cavity lifetime, and renormalizing the wavefunction at each step often removes the remaining differences.[20] In quantum optics, the steady state of a weakly driven cavity is pure to leading order in the drive,[74,75] and for some systems the whole master-equation evolution follows from the no-jump evolution.[76] To determine which observables the no-jump part gives for a correlated molecule whose ground state contains virtual photons, we solved the full Lindblad equation and the no-jump evolution exactly for $H_2$ with the parameters of Figs. 1 and 4 and compared their observables (Fig. 5 and Table 1).

The bare-mode collapse operator, $\hat{b}$ or the displaced $\hat{b} - \langle\hat{b}\rangle$, is the usual choice in Lindblad treatments of molecular polaritons,[17,18,20] but the correlated ground state is not stationary under it. The squared norm of $(\hat{b} - \langle\hat{b}\rangle)|\mathrm{G}\rangle$ is the incoherent photon number of the ground state, which the counter-rotating terms and the dipole self-energy make nonzero ($9.4 \times 10^{-4}$ here).[49,50] This dissipator lets the ground state emit virtual photons that a physical cavity does not emit,[52,53,77] although these photons remain in the ground state of a lossy cavity.[78] Without any excitation, the bare-mode dissipator raises the photon number of the ground state by about the same amount, whereas the dressed collapse operator leaves the ground state dark. Lindblad operators that act only on the photon mode also remain valid at ultrastrong coupling when the spectral density of the environment is suppressed at negative frequencies.[79]

Table 1: Maximum deviation (%) of the normalized no-jump evolution from the Lindblad dynamics with the dressed collapse operator for $H_2$/6-31G ($\lambda = 0.05$ a.u., $\kappa = 0.02$ a.u.) over 400 a.u. after the excitation. The no-jump evolution uses the bare-mode collapse operator, which the coupled-cluster propagation damps, or the dressed one, and starts from the same state as the Lindblad dynamics. Coherence deviations are those of $\langle \hat{b} \rangle$ relative to the kick $\varepsilon$, and population deviations are those of the photon number of the displaced mode, $\langle \hat{n}' \rangle$, relative to its excitation, which is $\varepsilon^2$ for the kicks. The drive row compares the periodic steady states under the drive of Fig. 4, averaged over one drive period.

| excitation | quantity | bare-mode | dressed |
|---|---|---|---|
| kick, $\varepsilon = 0.02$ | coherence | 0.06 | 0.006 |
| | population | 4.1 | 0.009 |
| kick, $\varepsilon = 0.2$ | coherence | 0.51 | 0.55 |
| | population | 0.14 | 0.19 |
| drive, $\eta = 10^{-3}$ | population | 0.47 | 0.78 |
| LP eigenstate | population | 115 | 115 |
| one-photon state | population | 86 | 86 |

Table 1 summarizes the comparison. In this $H_2$ model, the normalized no-jump evolution reproduces the coherence after coherent kicks and, provided the ground state is dark, the populations of coherently excited and driven states. With the bare-mode operator, the no-jump population after the small kick deviates in an initial transient that also occurs without a kick, because the correlated ground state is not stationary under this evolution. Started from the stationary state of $\hat{H}_{\text{eff}}$, as in the coupled-cluster propagation, the deviation becomes negligible (Supporting Information). With the dressed operator, the no-jump population follows the Lindblad population, so the jump term is negligible at this excitation. In the periodic steady state under the drive, the ratio $g_{\text{c}}^{(2)}$ of the period-averaged moments of Ref. 49 (Supporting Information) is nearly the same for the no-jump state and the Lindblad dynamics, whereas the bare-mode dissipator gives a value about 13% smaller. In contrast, the jump term is essential after incoherent excitation. Starting from the lower-polariton eigenstate, the Lindblad photon number decays almost to zero, whereas the normalized no-jump state keeps about half a photon (Fig. 5c), because the no-jump evolution cannot return population to the ground state.

QED-CCSD-22 reproduces the coherent part of these no-jump dynamics (Fig. 5a,d). After the small kick, its co-rotating polariton amplitudes agree with those of the exact no-jump coherence to within 1%, and under the drive its estimate of the excitation of $\langle \hat{n}' \rangle$, which is $n_{\text{coh}}$, agrees equally well with that of the Lindblad steady state. With $\hat{\Lambda}$ fixed, the moments respond to the kick only through $\hat{T}(t)$, whose linear response oscillates at the eigenvalues $\omega_k - i\gamma_k$ of the Jacobian. They therefore miss the counter-rotating component of $\langle \hat{b} \rangle$ after the kick, whose size relative to the co-rotating one, $|\langle \mathrm{G}|\hat{b}^\dagger|\mathrm{LP}\rangle|/|\langle \mathrm{G}|\hat{b}|\mathrm{LP}\rangle| = 0.03$, comes from the counter-rotating terms of Eq. (1). Propagating $\hat{\Lambda}$ with the complex frequency does not remove this limitation, because the left state then evolves with the inverse rather than the adjoint of the ket propagator, and the moments leave the normalized no-jump expectation values within about one photon lifetime (Supporting Information).

The comparison so far uses a drive in the linear regime. Under a stronger drive, the exact steady state saturates. At the LP, the amplitude per unit drive of the Lindblad dynamics falls below its linear value by 5% at $\eta = 3 \times 10^{-3}$, where the LP eigenstate holds a few percent of the population (Supporting Information). The normalized no-jump amplitude hardly changes, whereas the QED-CCSD-22 amplitude increases. Because $e^{-\hat{T}} \hat{b}\, e^{\hat{T}} = \hat{b} + [\hat{b}, \hat{T}]$, the coupled-cluster amplitude with a fixed $\hat{\Lambda}$ is linear in the amplitudes of $\hat{T}(t)$, which are normalized to the reference component of the state. Accordingly, we find that it follows the ratio $\langle G|\hat{b}|\psi\rangle/\langle G|\psi\rangle$ evaluated with the exact no-jump state, which grows as the ground-state weight of $|\psi\rangle$ decreases (Supporting Information). The complex-frequency propagation with a fixed $\hat{\Lambda}$ thus describes drives that leave the polaritons nearly unpopulated, below about 1% in the LP eigenstate for $H_2$, which includes weak-probe transmission and absorption spectra. Saturation and pumped populations require the jump term and a time-dependent $\hat{\Lambda}$.

## 5. Conclusions

In this work, cavity loss was added to real-time QED-CCSD-22 through a complex cavity frequency. The resulting no-jump theory gives polariton linewidths, which agree between the complex Jacobian and real-time propagation and are partitioned by the biorthogonal photon numbers of the polaritons. Electron correlation enters these linewidths through the polariton energies and photon numbers, whereas the second photon changes them by less than 0.2%. The theory also gives driven-dissipative steady states, whose lineshapes carry the same linewidths and show interference between the two polaritons.

A comparison with exact Lindblad dynamics of $H_2$ shows that the no-jump description reproduces coherences and the populations and statistics of weakly and coherently driven states when the dissipator leaves the correlated ground state dark. Population decay after incoherent excitation and the saturation of strongly driven polaritons require the jump term. Without the Lamb shift, cavity loss changes the pole splitting only at second order in $\kappa/g$, as in the two-level model, and leaves the transmission peaks of $H_2$ at their lossless positions, whereas the peak separation of the molecular response decreases at twice the rate of the pole splitting. A strong loss-induced reduction of the pole splitting thus requires $\kappa$ to be a sizable fraction of $2\Omega_0$, as for $H_2O$ at $Q \approx 15$, whereas for p-nitroaniline at $Q = 60$ the reduction is 0.05%. The time-averaged Hamiltonian of Ref. 36 describes a different quantity, whose change with loss is set by the averaging time.

Population decay after incoherent excitation and the response of saturated polaritons require a trace-preserving Lindblad formulation of QED-CC built on a dissipator that leaves the correlated ground state dark, together with a time-dependent $\hat{\Lambda}$. With both, the intracavity statistics of Ref. 49 could be extended to the $g^{(2)}(\tau)$ of the emitted light through dressed-state master equations and photodetection operators.[53,55,61,80]

### Data Availability Statement

The data that support the findings of this study are available within the article and its Supporting Information. An archival snapshot of the source code, working equations, input scripts, and archived outputs is deposited on Zenodo (DOI: 10.5281/zenodo.23601959).

### Acknowledgement

This work was supported by the Air Force Office of Scientific Research under AFOSR Award No. FA9550-24-1-0199. We acknowledge the use of artificial intelligence in writing some of the code used in this study and in revising and editing the text. The authors assume full responsibility for the correctness of the results.

### Supporting Information Available

Validation of each part of the implementation, including the real-time linewidths and the exact complex polariton roots; derivation of the two-photon residual metric; the complex stationary state; linewidth and steady-state extraction, including the fit sensitivity of the $H_2O$ linewidths; the Lindblad comparison with both collapse operators, the bath model, the periodic steady state under the drive, and the strong-drive scan; the $H_2O$ exceptional-point scan and the comparison with QED-CCSD-21 and QED-CIS; the comparison of splitting definitions, including the time-averaged Hamiltonian, the transmission spectra, the molecular response, and the $H_2O$ values; and the displacement of the damped mode for a polar orientation (PDF).

**Supporting Information: Cavity Loss in QED Coupled-Cluster Theory: Polariton Linewidths, Driven Steady States, and the Rabi Splitting**

Kurt R. Brorsen, Department of Chemistry, University of Missouri, Columbia, Missouri 65211, United States (brorsenk@missouri.edu)

## S1. Systems and propagation settings

The molecules, coupling strengths, and cavity frequencies are given in Sec. 3 of the main text. The integrator was also checked for $H_2$/STO-3G at $\lambda = 0.05$ a.u. and $\omega = 10$ eV. The QED-HF reference is that of Ref. 1. For $H_2$, QED-HF is converged to $10^{-10}$, and the QED-CCSD-22[2,3] amplitude and $\Lambda$ equations are converged to residual norms of $10^{-12}$ and $10^{-11}$, respectively. For $H_2O$, the corresponding thresholds are $10^{-9}$, $10^{-11}$, and $10^{-10}$. The complex stationary state (Sec. S4) is converged to $10^{-11}$ for both molecules.

Kicked trajectories start with a dipole kick in the space of single excitations or with a one-photon kick, each of strength $10^{-3}$. The propagation uses ETDRK4[4,5] with the diagonal of the Jacobian as the linear part. With cavity loss this diagonal acquires $-in\kappa/2$ on the rows with $n$ photons, so the $\varphi$ coefficients are evaluated by a contour integral over a full complex circle instead of the real-axis shortcut of Ref. 5. The Hermitian spectra use trajectories of 2000 a.u. The Jacobian is built about the stationary state by central differences with a step of $10^{-5}$, in a reduced basis with one mode per pair $(ai) \le (bj)$ of each block with two electronic excitations. This basis has 29 modes for $H_2$ and 2582 modes for $H_2O$. Real-valued differences suffice for complex $\omega$, because the equations are holomorphic in the amplitudes.

## S2. Validation

Table S1 lists the checks of the implementation against exact references.

Table S1: Deviations of the implementation from exact references. Energies are in hartree, and "rel" marks relative deviations.

| check | reference | deviation |
|---|---|---|
| *Stationarity and integrator order ($H_2$/STO-3G)* | | |
| deviation norm, 1000 steps, no kick | stationarity | $7.6 \times 10^{-17}$ |
| step-halving error ratio, kicked | fourth order (16) | 0.05 |
| *Jacobian eigenvalues at $\lambda = 0$* | | |
| excitation energies, $H_2$/6-31G | EOM-CCSD | $1.1 \times 10^{-11}$ |
| excitation energies, $H_2O$/6-31G | EOM-CCSD | $4.5 \times 10^{-8}$ |
| photon roots $\omega$, $2\omega$, $E_k + 2\omega$ | analytic | $\le 2.0 \times 10^{-12}$ |
| *Real-time spectra ($\lambda = 0.05$, $\kappa = 0$)* | | |
| $H_2$ and $H_2O$ polariton peaks | Jacobian doublet | $\le 3.6 \times 10^{-5}$ |
| *Cavity loss at $\lambda = 0$ ($\kappa = 0.02$)* | | |
| photon roots $\omega - i\kappa/2$, $2\omega - i\kappa$, $E_1 + 2(\omega - i\kappa/2)$ | analytic | $\le 7.6 \times 10^{-12}$ |
| coherent-photon-number decay $e^{-\kappa t}$ | analytic | $4.2 \times 10^{-15}$ rel |
| *Cavity-loss linewidths ($\lambda = 0.05$, $H_2$, $\kappa \le 0.02$)* | | |
| complex polariton roots, real part | exact $\hat{H}_{\text{eff}}$ diagonalization | $8.9 \times 10^{-8}$ |
| complex polariton roots, linewidth | exact $\hat{H}_{\text{eff}}$ diagonalization | $5.9 \times 10^{-7}$ rel |
| real-time linewidths, dipole signal | Jacobian eigenvalues | $\le 5.3 \times 10^{-4}$ rel |

| check | reference | deviation |
|---|---|---|
| real-time linewidths, photon displacement | Jacobian eigenvalues | $\leq 1.1 \times 10^{-3}$ rel |
| *Photon observables* | | |
| moment evaluation at the ground state | Hermitian implementation | $1.8 \times 10^{-22}$ |
| stationary $n_c$ and $g_c^{(2)}$ | exact lossy stationary state | 0.017% / 0.4% |
| *Continuous-wave drive ($H_2$, $\kappa = 0.02$)* | | |
| driven damped mode, $\Delta t = 0.1$ | analytic solution | $2.1 \times 10^{-10}$ |
| drive terms | dressed-integral construction | $2.8 \times 10^{-17}$ |
| steady lineshape pole position | Jacobian LP | $1.0 \times 10^{-5}$ |
| steady lineshape width | $2\gamma_{LP}$ | $1.6 \times 10^{-3}$ rel |

The stationarity check propagates the converged $H_2$/STO-3G ground state for 1000 ETDRK4 steps without a kick, and the integrator order is the ratio of the errors at $\Delta t = 0.1$ and 0.05 a.u. after a dipole kick. The Jacobian eigenvalues at $\lambda = 0$ are compared with EOM-CCSD roots from PySCF, whose Davidson convergence sets the $H_2O$ deviation, and the real-time peak positions are taken from Padé-accelerated Fourier transforms of the kicked dipole signal. The correlated complex roots are compared with the excitation energies $E_k - E_0$ from exact diagonalization of $\hat{H}_{\text{eff}} = \hat{H} - i(\kappa/2)\hat{b}^\dagger\hat{b}$ in the basis of the exact reference of Sec. S6, where $E_0$ is the complex eigenvalue of the stationary state. Their deviation is nearly independent of $\kappa$, so it comes from the truncation of the photon excitation rank rather than from the loss.

### Linewidths

Table S4 lists the complex-Jacobian linewidths of Fig. 1 of the main text. For $H_2$, the $\kappa \to 0$ slopes $\gamma_k/(\kappa/2)$ of the Jacobian about the complex stationary state, 0.5510 (LP) and 0.4506 (UP), equal the exact excess photon numbers $\langle k|\hat{n}|k\rangle - \langle \mathrm{G}|\hat{n}|\mathrm{G}\rangle$ of the reference of Sec. S6, whereas the Jacobian about the real-$\omega$ ground state gives the biorthogonal photon numbers. The squared one-photon weights of the normalized right eigenvectors are 0.704 (LP) and 0.611 (UP).

Table S2 compares the real-time linewidths with the Jacobian. The photon-displacement widths of $H_2$ are slightly high because of the time-step error of the integrator, which halving the time step largely removes. For $H_2O$ at $\kappa = 0.01$ a.u., the polariton doublet is weak compared with nearby electronic peaks of the dipole spectrum, so the fitted widths depend on the number of fitted modes and on the time window (Table S3). The fits with ten modes agree with the Jacobian widths to within 2%.

Table S2: Real-time linewidths of $H_2$/6-31G versus complex-Jacobian eigenvalues at $\kappa = 0.02$ a.u. and $\lambda = 0.05$ a.u., starting from the complex stationary state. The real-time values are variable-projection fits of the kicked dipole signal and of the coherent photon displacement $\mathrm{Re}\,\delta M_{\hat{b}}$ (photon channel). Frequencies and linewidths are in hartree.

| system / channel | mode | $\omega_{\text{RT}}$ | $\mathrm{Re}\,\omega_{\text{Jac}}$ | $\gamma_{\text{RT}}$ | $-\mathrm{Im}\,\omega_{\text{Jac}}$ |
|---|---|---|---|---|---|
| $H_2$, dipole | LP | 0.530758 | 0.530754 | $5.5152 \times 10^{-3}$ | $5.5154 \times 10^{-3}$ |
| | UP | 0.599129 | 0.599121 | $4.5030 \times 10^{-3}$ | $4.5006 \times 10^{-3}$ |
| $H_2$, photon | LP | 0.530759 | 0.530754 | $5.5203 \times 10^{-3}$ | $5.5154 \times 10^{-3}$ |
| | UP | 0.599128 | 0.599121 | $4.5055 \times 10^{-3}$ | $4.5006 \times 10^{-3}$ |

Table S3: Real-time linewidths of $H_2O$/6-31G at $\kappa = 0.01$ a.u. ($\lambda = 0.05$ a.u.) from variable-projection fits of the kicked dipole signal with different numbers of damped cosines and time windows, as relative deviations from the complex-Jacobian values $\gamma_{\mathrm{LP}} = 2.9586 \times 10^{-3}$ and $\gamma_{\mathrm{UP}} = 2.0421 \times 10^{-3}$ a.u. The fitted pole positions agree with the Jacobian to $3.3 \times 10^{-5}$ Ha or better.

| modes | window (a.u.) | LP width | UP width |
|---|---|---|---|
| 7 | 0–1600 | +4.9% | −2.5% |
| 10 | 0–1600 | −0.17% | +0.41% |
| 10 | 0–1200 | −0.03% | +0.36% |
| 10 | 0–800 | −1.6% | −0.35% |

Table S4: Complex-Jacobian linewidths of Fig. 1 of the main text ($\lambda = 0.05$ a.u., cavity in resonance with the $\lambda = 0$ excitation) and their relative deviation ("rel") from the first-order partition $(\kappa/2)\, n_k$, with $n_{\mathrm{LP}} = 0.5516$ and $n_{\mathrm{UP}} = 0.4490$ for $H_2$ and 0.5803 and 0.4198 for $H_2O$. For $H_2O$, the doublet splitting is compared with the resonant two-level value $\sqrt{\Omega_0^2 - (\kappa/2)^2}$, with $\Omega_0 = 1.0166 \times 10^{-2}$ Ha. Linewidths and splittings are in $10^{-3}$ a.u. The sum $\gamma_{\mathrm{LP}} + \gamma_{\mathrm{UP}}$ deviates from $\kappa/2$ by $1.6 \times 10^{-3}$ ($H_2$) and $1.5 \times 10^{-4}$ ($H_2O$) relative.

| | $\kappa$ (a.u.) | $\gamma_{\mathrm{LP}}$ | rel | $\gamma_{\mathrm{UP}}$ | rel | splitting | resonant |
|---|---|---|---|---|---|---|---|
| $H_2$ | 0.005 | 1.3776 | $1.0 \times 10^{-3}$ | 1.1264 | $3.4 \times 10^{-3}$ | | |
| | 0.01 | 2.7557 | $8.2 \times 10^{-4}$ | 2.2523 | $3.1 \times 10^{-3}$ | | |
| | 0.02 | 5.5154 | $9.6 \times 10^{-5}$ | 4.5006 | $2.3 \times 10^{-3}$ | | |
| $H_2O$ | 0.0025 | 0.72613 | $1.1 \times 10^{-3}$ | 0.52406 | $1.2 \times 10^{-3}$ | 10.091 | 10.089 |
| | 0.005 | 1.4569 | $4.3 \times 10^{-3}$ | 1.0434 | $5.7 \times 10^{-3}$ | 9.8630 | 9.8542 |
| | 0.01 | 2.9586 | $2.0 \times 10^{-2}$ | 2.0421 | $2.7 \times 10^{-2}$ | 8.8999 | 8.8518 |
| | 0.02 | 6.9176 | $1.9 \times 10^{-1}$ | 3.0840 | $2.7 \times 10^{-1}$ | 4.2553 | 1.8317 |

### Photon observables and the drive

For the complex stationary state, QED-CCSD-22 with the lossless ground-state $\hat{\Lambda}_0$ gives the biorthogonal incoherent photon number $n_{\mathrm{c}} = 9.43 \times 10^{-4}$, with a small imaginary part, and $g_{\mathrm{c}}^{(2)} = 5.81$. The exact reference is the least-damped eigenvector of $\hat{H}_{\mathrm{eff}}$ with the collapse operator $\hat{b} - \langle \hat{b} \rangle$ (Sec. S6) in the biorthogonal pairing of the coupled-cluster construction, and the deviations in Table S1 are consistent with $\hat{\Lambda}_0$, which is correct only to $O(\kappa)$. The imaginary part of the complex stationary energy (Sec. S4) matches the exact value for the displaced mode, $-9.43 \times 10^{-6}$ Ha, and not that for $\hat{b}$, which is more than five times larger, so the propagation damps the displaced mode.

The drive terms of Sec. 2.3 of the main text agree with a construction that absorbs the drive into the one-electron integrals through $\hat{d} \to \hat{d} + \eta\, \hat{1}/N_{\mathrm{el}}$, for doubles symmetrized under $(ai) \leftrightarrow (bj)$, the space that the propagation reaches.

## S3. The two-photon residual metric

The symbolically generated residuals $R_\mu$ of the ground-state solver differ from the projections $\Omega_\mu$ of Eq. 2 of the main text on the rows with two photons,

$$R_\mu \;=\; W_\mu \, \Omega_\mu, \tag{1}$$

$$W_\mu = 1/2 \quad \text{for} \quad \mu \in \{p_2,\, s_1^2,\, s_2^2\},$$

$$W_\mu = 1 \quad \text{otherwise},$$

so the propagation uses $\Omega = W^{-1}R$. The two-photon amplitudes are stored with the photon Fock factor $1/2!$. The excitation operator $\hat{\tau}_\mu = 1/2\,\hat{A}_\mu(\hat{b}^\dagger)^2$ attached to such an amplitude produces $\hat{\tau}_\mu|\mathrm{HF}\rangle = |\phi_\mu\rangle/\sqrt{2!}$, where $|\phi_\mu\rangle$ is the normalized two-photon state and $\hat{A}_\mu$ the accompanying electronic operator, so the biorthogonal bra is $\langle\mu| = \sqrt{2!}\,\langle\phi_\mu|$, whereas the symbolic generator returns $\langle\phi_\mu|\bar{H}|\mathrm{HF}\rangle/\sqrt{2!}$. The two factors combine to $\Omega_\mu = 2!\,R_\mu$. With the metric, the Jacobian for $\lambda \to 0$ places the photon roots at $\omega$ and $2\omega$ and the mixed roots at $E_k + 2\omega$, and with loss it places the two-photon root at $2\omega - i\kappa$ (Table S1). A ground-state solver cannot detect the factor, because $R = 0$ and $W^{-1}R = 0$ have the same solutions, but without it the propagation evolves the two-photon amplitudes at half their frequencies.

### S4. The complex stationary state

With $\omega \to \omega - i\kappa/2$, the Hermitian ground state is no longer stationary. The residual appears only on the rows with photons, with a norm of the order of $\kappa/2$ times that of the photon amplitudes. The complex stationary state is found by a quasi-Newton iteration from the real-$\omega$ ground state, with the complex Jacobi diagonal as preconditioner and DIIS over the eight most recent iterates, to a residual norm below $10^{-11}$, which takes about ten iterations. For $H_2$ at $\kappa = 0.02$ a.u., the complex correlation energy is $E_{\mathrm{corr}} = -0.0264718898 - 9.427 \times 10^{-6}\,i$ Ha, compared with $-0.0264719741$ Ha at real $\omega$.

All kicked and driven trajectories start from the complex stationary state, because a propagation from the real-$\omega$ ground state relaxes onto it through a photon transient (about 40% of the kick scale) that does not appear in the dipole signal but does appear in photon observables.

### S5. Steady-state amplitude and linewidth extraction

Linewidths of kicked trajectories are obtained by a variable-projection fit of the real signal to damped cosines,

$$s(t) \;=\; c_0 + \sum_k e^{-\gamma_k t}\,[a_k \cos(\omega_k t) + b_k \sin(\omega_k t)], \qquad (2)$$

where each $\gamma_k$ is the half width at half maximum of a Lorentzian line and corresponds to $-\mathrm{Im}$ of a Jacobian eigenvalue. The linear coefficients $(c_0, a_k, b_k)$ are solved exactly at every iteration, and the nonlinear search runs over $(\omega_k, \gamma_k)$ within $\pm 0.02$ a.u. of seed frequencies from a damped fast Fourier transform, with $\gamma_k \le 1$ a.u. For $H_2O$, the polariton doublet reaches only about 7% of the tallest peak of the damped transform, so it is seeded by a separate search over 0.28–0.34 Ha, and up to five electronic peaks are added from the full window. The ten-mode fits of Table S3 add three electronic modes at the largest remaining peaks (1.684, 1.695, and 1.928 Ha).

Driven steady states ($T = 500$ a.u. per drive frequency) are analyzed by a linear least-squares fit of the recorded one-photon amplitude $p_1$,

$$s(t) \;=\; A\,e^{-i\omega_d t} + B\,e^{+i\omega_d t} + \sum_k C_k\,e^{-iz_k t}, \tag{3}$$

where $A$ and $B$ are the co- and counter-rotating steady amplitudes and the switch-on transient oscillates at the known complex Jacobian poles $z_k$. Across the scan of Fig. 4 of the main text, $|B|$ stays close to its analytic value $\eta/4\omega = 4.4\times10^{-4}$. The lineshape is fitted to

$$A(\omega_d) \;=\; \frac{r}{\omega_0-\omega_d-i\gamma} + \frac{r_{\mathrm{UP}}}{\mathrm{Re}\,z_{\mathrm{UP}}-\omega_d+i\,\mathrm{Im}\,z_{\mathrm{UP}}} \;+\; c_0, \tag{4}$$

with the UP pole fixed at its Jacobian value, the residues and $c_0$ solved linearly, and a simplex search over $(\omega_0,\gamma)$. The fitted pole agrees with the Jacobian LP (Table S1). The pole width $2\gamma$ is the full width at half maximum of the Lorentzian $|A|^2$ of an isolated pole, whereas $|A(\omega_d)|$ of an isolated pole has a width of $2\sqrt{3}\,\gamma$, and the UP pole changes its shape further.

## S6. Comparison with exact Lindblad dynamics

The exact reference of Sec. 3 of the main text uses the lab frame and the electronic dipole convention of the coupled-cluster calculations. The Lindblad equation, Eq. 3 of the main text with the collapse operator written here as $\hat{c}$, is integrated at $\Delta t = 0.02$ a.u. in the eigenstates of $\hat{H}$ that are connected to the ground state by $\hat{b}+\hat{b}^\dagger$ and lie less than $6\omega$ above it (38 states for $n_{\max}=12$).

The bare-mode collapse operator is the displaced mode $\hat{c}=\hat{b}-\langle\hat{b}\rangle_{\mathrm{G}}$, which the coupled-cluster propagation damps for $H_2$. A dissipator built on $\hat{b}$ instead adds a Hermitian drive of order $\kappa\langle\hat{b}\rangle_{\mathrm{G}}$, which depends on the origin and changes the coherence after a kick by up to 2% of its scale. The dressed collapse operator is

$$\hat{c} \;=\; \sum_{E_k<E_j}\sqrt{\frac{E_j-E_k}{\omega}}\;\langle k|\hat{b}+\hat{b}^\dagger|j\rangle\;|k\rangle\langle j|, \tag{5}$$

where $|j\rangle$ are the eigenstates of $\hat{H}$ with energies $E_j$. It equals the jump operator of the universal Lindblad equation,[6,7]

$$\hat{L}_{\mathrm{ULE}} \;=\; \sum_{m,n}\sqrt{2\pi J(E_n-E_m)}\;\langle m|\hat{b}+\hat{b}^\dagger|n\rangle\;|m\rangle\langle n|, \tag{6}$$

divided by $\sqrt{\kappa}$, for an ohmic bath at zero temperature, $2\pi J(\nu)=\kappa\nu/\omega$ for $\nu>0$ and $J(\nu)=0$ otherwise, where $J$ includes the system–bath coupling. The Lamb shift is omitted. The operator reduces to $\hat{b}$ at $\lambda=0$ and annihilates the correlated ground state exactly.

Because $\hat{c}$ is a single operator, the dissipator keeps the cross terms between transitions. For the two polaritons, which decay into the same ground state, $\langle\mathrm{LP}|\hat{c}^\dagger\hat{c}|\mathrm{UP}\rangle = -0.497$ is close to $-(0.550\times0.450)^{1/2}$, so the two decay channels interfere almost completely, and this element alone produces the loss dependence of the pole splitting in Table S9. A secular master equation with one collapse operator per transition drops it, because its $\hat{H}_{\mathrm{eff}}$ is diagonal in the eigenstates of $\hat{H}$, and leaves the real parts of the poles at their lossless values.

The no-jump evolution under $\hat{H}_{\text{eff}} = \hat{H} - i(\kappa/2)\hat{c}^\dagger\hat{c}$ is reported in normalized form, $\langle\hat{O}\rangle = \langle\psi|\hat{O}|\psi\rangle/\langle\psi|\psi\rangle$. As in Ref. 8, $n_\text{c} = \langle\delta\hat{b}^\dagger\delta\hat{b}\rangle$ and $G_\text{c} = \langle\delta\hat{b}^{\dagger 2}\delta\hat{b}^2\rangle$ with $\delta\hat{b} = \hat{b} - \langle\hat{b}\rangle(t)$, and $g_\text{c}^{(2)} = G_\text{c}/{n_\text{c}}^2$. For the ground state, $\langle\hat{n}'\rangle = n_\text{c} = 9.43 \times 10^{-4}$ and $g_\text{c}^{(2)} = 5.83$.

Under the bare-mode dissipator, the ground state is not stationary, because $\|\hat{c}|\text{G}\rangle\|^2 = n_\text{c}$. Without any excitation, the Lindblad photon number rises by $9.6 \times 10^{-4}$ by $T = 3000$ a.u., so that the incoherent photon number doubles. The normalized no-jump evolution with this operator, started from $|\text{G}\rangle$, changes $\langle\hat{n}'\rangle$ within 2 a.u. by 4.1% of the excitation $\varepsilon^2$ of the small kick, which is the bare-mode population entry for $\varepsilon = 0.02$ in Table 1 of the main text. Started from the stationary state of $\hat{H}_{\text{eff}}$, it deviates from the dressed Lindblad population by at most 0.09% of $\varepsilon^2$. The dressed operator gives the linewidths $\gamma_k/(\kappa/2) = 0.5507$ (LP) and 0.4493 (UP) at $\kappa = 0.02$ a.u., compared with 0.5515 and 0.4501 for the bare-mode operator, whereas a flat bath spectral density in Eq. (5) gives 0.584 and 0.422.

Table 1 of the main text starts from coherent kicks $\hat{D}(\varepsilon)|\text{G}\rangle$ with $\varepsilon = 0.2$ and 0.02, followed to $T = 400$ a.u. ($\kappa T = 8$), from the exact lower-polariton eigenstate or the normalized one-photon state $\hat{b}^\dagger|\text{G}\rangle$, or uses the drive $\eta\cos(\omega_d t)(\hat{b} + \hat{b}^\dagger)$ at the LP frequency with $\eta = 10^{-3}$. The periodic steady state under the drive, $\hat{\rho}(t) = \sum_m \hat{\rho}_m\, e^{-im\omega_d t}$, follows from the harmonic balance $(\mathcal{L}_0 + im\omega_d)\hat{\rho}_m + (\eta/2)\mathcal{V}(\hat{\rho}_{m-1} + \hat{\rho}_{m+1}) = 0$ with $\mathcal{V}\hat{\rho} = -i[\hat{b} + \hat{b}^\dagger, \hat{\rho}]$, solved by matrix continued fractions with $|m| \leq 6$. The normalized no-jump steady state is the Floquet state of $\hat{H}_{\text{eff}} + \eta\cos(\omega_d t)(\hat{b} + \hat{b}^\dagger)$ with the slowest decay. Photon numbers are averages over one drive period, and $g_\text{c}^{(2)}$ is the period average of $G_\text{c}$ divided by the square of the period average of $n_\text{c}$ (Table S5).

Table S5: Periodic steady state under the drive of the main text ($H_2$/6-31G, $\lambda = 0.05$ a.u., $\kappa = 0.02$ a.u., $\eta = 10^{-3}$, $\omega_d = 0.530754$ Ha). Photon numbers are averages over one drive period, $\langle\hat{n}'\rangle$ is the photon number of the displaced mode, and $|A|$ is the co-rotating amplitude of $\langle\hat{b}\rangle$. The ground-state value of $\langle\hat{n}'\rangle$ is $9.428 \times 10^{-4}$. For QED-CCSD-22, $\langle\hat{n}'\rangle$ is the coupled-cluster ground-state value, $9.427 \times 10^{-4}$, plus $n_{\text{coh}}$.

| | $\langle\hat{n}'\rangle$ ($10^{-3}$) | $n_\text{c}$ ($10^{-4}$) | $g_\text{c}^{(2)}$ | $\|A\|/\eta$ |
|---|---|---|---|---|
| Lindblad, dressed | 3.5745 | 9.577 | 10.27 | 51.13 |
| Lindblad, bare-mode | 4.5295 | 19.34 | 8.89 | 50.92 |
| no-jump, bare-mode | 3.5867 | 9.500 | 10.45 | 51.33 |
| no-jump, dressed | 3.5949 | 9.501 | 10.48 | 51.40 |
| QED-CCSD-22 | 3.5949 | — | — | 51.50 |

After the kick $\varepsilon = 0.02$, the exact no-jump coherence contains counter-rotating terms $e^{+i\omega_k^* t}$ with about 3% of the amplitude of the co-rotating LP term, in agreement with the ratio $|\langle\text{G}|\hat{b}^\dagger|\text{LP}\rangle|/|\langle\text{G}|\hat{b}|\text{LP}\rangle| = 0.0295$. The coupled-cluster moments with $\hat{\Lambda}_0$ contain the co-rotating terms, within 0.1% (LP) and 0.4% (UP) of the exact amplitudes, and no counter-rotating terms.

A time-dependent $\hat{\Lambda}$ does not give the normalized no-jump expectation values either. With a complex frequency, the left state evolves as $\langle\tilde{\Psi}_0|e^{+i\hat{H}_{\text{eff}}t}$, the inverse rather than the adjoint of the ket propagator, so the moments evaluate $\langle\tilde{\Psi}_0|e^{i\hat{H}_{\text{eff}}t}\hat{O}e^{-i\hat{H}_{\text{eff}}t}|\Psi_0\rangle$, whose cross terms grow as $e^{(\gamma_j - \gamma_k)t}$. In the exact model, this pairing departs from the normalized no-jump coherence by more than 1% at $\kappa t = 0.72$ and from the photon number at $\kappa t = 0.96$.

### Strong drive

Table S6 gives the steady state at the LP frequency for increasing drive strength. The exact periodic steady states use $n_{\max} = 12$, the states below $5\omega$, and $|m| \leq 6$, which are converged up to $\eta = 10^{-2}$. The no-jump values belong to the Floquet state connected to the ground state, which decays most slowly up to $\eta = 7 \times 10^{-3}$. The coupled-cluster amplitudes are fits of $\delta M_{\hat{b}}(t)$ over the last 500 a.u. of 4000-a.u. trajectories to the harmonics $0, \pm 1, \pm 2$ of $\omega_d$.

Table S6: Steady state at the LP frequency for increasing drive strength $\eta$ ($H_2$/6-31G, $\lambda = 0.05$ a.u., $\kappa = 0.02$ a.u.). Listed are the population of the exact LP eigenstate in the Lindblad steady state with the dressed collapse operator and the co-rotating amplitude of $\langle \hat{b} \rangle$ per unit drive, $|A|/\eta$, for that Lindblad steady state, for the normalized no-jump state with the bare-mode collapse operator, for the ratio $\langle \mathrm{G}|\hat{b}|\psi\rangle / \langle \mathrm{G}|\psi\rangle$ on the same no-jump state, and for QED-CCSD-22.

| $\eta$ | LP population | Lindblad | no-jump | $\langle \mathrm{G}\|\hat{b}\|\psi\rangle / \langle \mathrm{G}\|\psi\rangle$ | QED-CCSD-22 |
|---|---|---|---|---|---|
| $10^{-4}$ | $< 10^{-4}$ | 51.42 | 51.34 | 51.34 | 51.37 |
| $10^{-3}$ | 0.005 | 51.13 | 51.33 | 51.47 | 51.50 |
| $2 \times 10^{-3}$ | 0.019 | 50.29 | 51.27 | 51.87 | 51.89 |
| $3 \times 10^{-3}$ | 0.041 | 48.98 | 51.15 | 52.56 | 52.56 |
| $5 \times 10^{-3}$ | 0.101 | 45.36 | 50.41 | 54.86 | 54.65 |
| $10^{-2}$ | 0.259 | 35.42 | — | — | — |

### Displacement of the damped mode for a polar orientation

With $\hat{b}$ the photon operator of the untransformed Hamiltonian, the coupled-cluster propagation damps $\hat{b} - z_{\mathrm{HF}}$, with $z_{\mathrm{HF}} = \langle \boldsymbol{\lambda} \cdot \hat{\boldsymbol{\mu}} \rangle_{\text{QED-HF}} / \sqrt{2\omega}$, whereas the exact ground state is displaced by $\langle \hat{b} \rangle_{\mathrm{G}}$. For $\lambda = 0.05$ a.u. along the $C_2$ axis of $H_2O$, with the cavity in resonance with the lowest excitation polarized along this axis, the exact STO-3G reference gives $\delta = \langle \hat{b} \rangle_{\mathrm{G}} - z_{\mathrm{HF}} = 2.0 \times 10^{-3}$. Damping $\hat{b} - z_{\mathrm{HF}}$ then makes the imaginary part of the stationary energy more negative by $(\kappa/2)\,\delta^2$, or 1.3%, and broadens the linewidths by less than $2 \times 10^{-4}$ relative. In 6-31G, QED-CCSD-22 gives $\delta = 2.6 \times 10^{-3}$, a 1.0% change, and less than $5 \times 10^{-4}$ at $\kappa = 0.02$ a.u.

### S7. The exceptional-point loss scan

The $H_2O$/6-31G loss scan of Fig. 3 of the main text uses twenty loss rates from 0.002 to 0.03 a.u. At each, the complex stationary state of Sec. S4 is converged to $10^{-11}$, and the complex Jacobian (2582 modes) is built and diagonalized. The LP and UP roots are followed by continuation in $\kappa$, with each root assigned to the nearest complex eigenvalue of the previous step, because the assignment by photon weight loses its meaning where the eigenvectors mix. The roots at each loss rate are in the archived data.

The detuned two-level model of the main text uses $\Delta = (n_{\mathrm{LP}} - n_{\mathrm{UP}})\,\Omega_0 = 1.6320 \times 10^{-3}$ Ha and $g = \sqrt{\Omega_0^2 - \Delta^2}/2 = 5.0173 \times 10^{-3}$ Ha, with the cavity below the dressed excitation, as the larger LP photon weight requires. It reproduces the splittings and linewidths of the twenty scanned roots to 0.21% or better, and its eigenvalues coalesce at the complex loss rate $\kappa = 0.02007 + 0.00326\,i$ a.u. The detuning is the dipole self-energy shift of the excitation relative to the cavity frequency, which is set in resonance with the excitation energy at $\lambda = 0$. For $H_2$, exact diagonalization of the reference Hamiltonian of Sec. S6 without the bilinear coupling places the dressed excitation $6.5 \times 10^{-3}$ Ha above $\omega$, close to the $\Delta$ of Table S7.

### Lower-level methods

QED-CCSD-21 keeps the electronic single and double excitations and one photon (amplitudes $t_1$, $t_2$, $p_1$, $s_1^1$, $s_2^1$), and its complex roots were computed as those of QED-CCSD-22. The non-Hermitian QED-CIS of Refs. 9 and 10 diagonalizes the Hamiltonian in the space of the QED-HF determinant and the singly excited singlet configurations, each with zero or one photon, in the coherent-state basis. We implemented it on the QED-HF of this work and checked it against the exact Pauli–Fierz Hamiltonian projected onto the same configurations. As in the coupled-cluster calculations, the complex frequency enters only the photon energy, and the linewidths are taken from the complex excitation energies $E_k - E_0$. The $\lambda = 0$ CIS excitation of $H_2O$ lies 0.038 Ha above EOM-CCSD, which would leave the QED-CIS doublet strongly detuned at the cavity frequency of the main text, so Table S7 uses the CIS excitation energy as the cavity frequency. The complex roots of each method follow its own detuned two-level model to 0.2%. With the complex bilinear prefactor $\sqrt{\tilde{\omega}/2}$ of Ref. 9, the QED-CIS ground root of $H_2$ at $\kappa = 0.02$ a.u. acquires a positive imaginary energy, $+1.2 \times 10^{-5}$ Ha instead of $-1.1 \times 10^{-5}$ Ha, and the $H_2$ linewidths at $\kappa = 0.01$ a.u. change by $-6\%$ (LP) and $+7\%$ (UP).

Table S7: Polariton doublets with lower-level methods ($\lambda = 0.05$ a.u., cavity in resonance with each method's $\lambda = 0$ excitation). $\Omega_0$ is the lossless splitting, $\Delta = (n_{\mathrm{LP}} - n_{\mathrm{UP}})\Omega_0$ the detuning, $\gamma_k$ the linewidths at $\kappa = 0.01$ a.u., "split" the real splitting at $\kappa = 0.02$ a.u., and $\kappa_c$ the loss rate of closest approach of the detuned two-level model. Energies and rates in $10^{-3}$ a.u. except $\omega$ (hartree).

| | method | $\omega$ | $\Omega_0$ | $\Delta$ | $n_{\mathrm{LP}}$ | $\gamma_{\mathrm{LP}}$ | $\gamma_{\mathrm{UP}}$ | split | $\kappa_c$ |
|---|---|---|---|---|---|---|---|---|---|
| $H_2O$ | QED-CCSD-22 | 0.3082 | 10.17 | 1.63 | 0.580 | 2.959 | 2.042 | 4.255 | 19.8 |
| | QED-CCSD-21 | 0.3082 | 10.16 | 1.63 | 0.580 | 2.959 | 2.042 | 4.247 | 19.8 |
| | QED-CIS | 0.3462 | 10.65 | 1.43 | 0.567 | 2.877 | 2.122 | 4.747 | 20.9 |
| $H_2$ | QED-CCSD-22 | 0.5626 | 69.09 | 7.08 | 0.552 | 2.756 | 2.252 | — | — |
| | QED-CIS | 0.5603 | 76.15 | 6.00 | 0.540 | 2.693 | 2.298 | — | — |

## S8. Cavity loss and the Rabi splitting

All calculations in this section use $H_2$/6-31G with the parameters of the main text. Relative changes refer to the lossless splitting at the same cavity frequency, $\Omega_0 = 0.0690870$ Ha for the exact reference.

### The time-averaged Hamiltonian

Bruno et al. obtain a Hermitian Hamiltonian in which the bilinear coupling is multiplied by $1 - \omega\tau/4Q$ and the dipole self-energy by $1 - \omega\tau/2Q$ (Eq. 14 of Ref. 11). With $\kappa = \omega/Q$ and $x = \kappa\tau/4$, this Hamiltonian reads

$$\hat{H}_x \; = \; \hat{H}_{\mathrm{el}} + \omega\, \hat{b}^\dagger \hat{b} - \sqrt{\omega/2}\; (1-x)\, (\boldsymbol{\lambda}\cdot\hat{\boldsymbol{\mu}})\, (\hat{b}^\dagger + \hat{b}) + 1/2\, (1-2x)\, (\boldsymbol{\lambda}\cdot\hat{\boldsymbol{\mu}})^2. \tag{7}$$

We implemented Eq. (7) exactly, with both prefactors, rather than as a rescaling of $\lambda$, which agrees with it only to first order in $x$. In QED-HF, the coherent-state shift that removes the mean-field part of the bilinear term is $z = (1-x)\langle\boldsymbol{\lambda}\cdot\hat{\boldsymbol{\mu}}\rangle/\sqrt{2\omega}$, and the electronic Hamiltonian becomes

$$\hat{H}_{\mathrm{el}} + 1/2\,(1-2x)\,(\boldsymbol{\lambda}\cdot\hat{\boldsymbol{\mu}})^2 - (1-x)^2\,\langle\boldsymbol{\lambda}\cdot\hat{\boldsymbol{\mu}}\rangle\,(\boldsymbol{\lambda}\cdot\hat{\boldsymbol{\mu}}) + 1/2\,(1-x)^2\,\langle\boldsymbol{\lambda}\cdot\hat{\boldsymbol{\mu}}\rangle^2, \tag{8}$$

while the bilinear term is $-\sqrt{\omega/2}\,(1-x)\,(\boldsymbol{\lambda}\cdot\hat{\boldsymbol{d}})(\hat{b}^\dagger+\hat{b})$. In the exact reference the same two prefactors multiply the bilinear and dipole self-energy matrices, and QED-CCSD-22 reproduces the exact polariton gaps of Eq. (7) at every $x$ (Table S8).

Three choices of the averaging time are compared. The first, $\kappa\tau = 0.688$, is the value at 5.430 eV, near the resonance of p-nitroaniline with Eq. (7), for the averaging time of Ref. 11 at $Q = 60$, $\tau = 5.01$ fs, which equals $Q/\omega$ at their highest cavity frequency of 7.89 eV. The second, $\tau = Q/\omega$, is the upper end of their averaging interval and gives $\kappa\tau = 1$ for every $Q$. The third is a fixed $\tau = 20/\omega$, the photon lifetime of a $Q = 20$ cavity, in analogy to their Fig. 3, which gives $\kappa\tau = 20/Q$ and is used for $Q \geq 20$ only, so that $\tau \leq Q/\omega$. Table S8 lists the resulting gaps.

Table S8: Relative change of the polariton gap of the time-averaged Hamiltonian, Eq. (7), for $H_2$/6-31G ($\lambda = 0.05$ a.u., $\omega = 0.562595$ Ha) from QED-CCSD-22 and from exact diagonalization, with $x = \kappa\tau/4$.

| averaging time | $\kappa\tau$ | QED-CCSD-22 | exact |
|---|---|---|---|
| Ref. 11 at $Q = 60$, any $Q$ | 0.688 | −17.32% | −17.32 % |
| $\tau = Q/\omega$, any $Q$ | 1 | −25.15% | −25.15 % |
| $\tau = 20/\omega$, $Q = 20$ | 1 | −25.15% | −25.15 % |
| $Q = 28$ | 0.714 | −17.97% | −17.97 % |
| $Q = 40$ | 0.5 | −12.58% | −12.58 % |
| $Q = 60$ | 0.333 | −8.39% | −8.39% |
| $Q = 100$ | 0.2 | −5.04% | −5.04% |
| $Q = 200$ | 0.1 | −2.52% | −2.52% |
| $Q = 500$ | 0.04 | −1.01% | −1.01% |
| $Q = 1000$ | 0.02 | −0.504% | −0.504 % |

**Pole splitting**

The complex QED-CCSD-22 polariton roots were computed at twelve loss rates from $Q = 1000$ to 5, with four intermediate loss rates between neighboring points for the continuation. The exact roots are eigenvalues of $\hat{H}_{\text{eff}}$ with the bare-mode collapse operator $\hat{b} - \langle\hat{b}\rangle$, which for $H_2$ is the operator that the propagation damps (Sec. S6), and with the dressed collapse operator of Eq. (5). QED-CCSD-22 agrees with the bare-mode roots to $1.4 \times 10^{-7}$ Ha, and the dressed collapse operator changes the pole splitting by at most 0.03 percentage points. The detuned two-level model of Sec. S7, built from the $H_2$ values of $\Omega_0$ and $\Delta$ in Table S7, follows the exact pole splitting closely (Table S9).

Table S9: Relative change of the pole splitting and of the peak separations of $H_2$/6-31G versus the quality factor ($\lambda = 0.05$ a.u.), with the polariton linewidths $\gamma_k = -\mathrm{Im}\,\omega_k$ from QED-CCSD-22. "Pole" is the separation of the real parts of the two complex roots (QED-CCSD-22, and exact with the dressed collapse operator); "transmission" is the peak separation of the exact transmission spectrum; "molecule" is the peak separation of the exact molecular response, which has a single maximum at $Q = 5$; "two-level" is the detuned two-level model.

| $Q$ | pole, CC | pole, exact | transmission | molecule | two-level | $\gamma_{\mathrm{LP}}$ $(10^{-3})$ | $\gamma_{\mathrm{UP}}$ $(10^{-3})$ |
|---|---|---|---|---|---|---|---|
| 1000 | −0.0008% | −0.0008% | 0.0000% | −0.0016% | −0.0008% | 0.155 | 0.127 |
| 500 | −0.0033% | −0.0033% | 0.0000% | −0.0065% | −0.0033% | 0.310 | 0.253 |
| 200 | −0.021% | −0.021% | 0.0000% | −0.041% | −0.021% | 0.775 | 0.634 |
| 100 | −0.082% | −0.082% | +0.0002% | −0.164% | −0.082% | 1.55 | 1.27 |
| 60 | −0.228% | −0.228% | +0.0004% | −0.455% | −0.228% | 2.58 | 2.11 |
| 40 | −0.514% | −0.514% | +0.0010% | −1.03% | −0.514% | 3.88 | 3.17 |
| 28 | −1.05% | −1.05% | +0.002% | −2.11% | −1.05% | 5.54 | 4.52 |
| 20 | −2.07% | −2.07% | +0.004% | −4.17% | −2.07% | 7.76 | 6.32 |
| 14 | −4.28% | −4.27% | +0.008% | −8.72% | −4.27% | 11.1 | 9.01 |
| 10 | −8.56% | −8.55% | +0.014% | −18.0% | −8.55% | 15.6 | 12.5 |
| 7 | −18.4% | −18.3% | +0.026% | −43.3% | −18.3% | 22.6 | 17.7 |
| 5 | −40.4% | −40.4% | +0.042% | one peak | −40.3% | 32.9 | 23.4 |

**Transmission spectrum**

The transmission spectrum is the weak-drive response of the cavity output in the Lindblad model with the dressed collapse operator of Eq. (5), for which the output field is proportional to $\langle\hat{c}\rangle$.[12] We assume a weak coherent input that drives the cavity mode, a single output port whose field is $\sqrt{\kappa}\,\hat{c}$, and no direct transmission of the input past the cavity, so the spectrum is the cavity output spectrum. The drive $\eta\cos(\omega_d t)\,\hat{X}$ with $\hat{X} = \hat{b} + \hat{b}^\dagger$ acts on the ground state $|\mathrm{G}\rangle$, which is stationary because $\hat{c}|\mathrm{G}\rangle = 0$. To first order in $\eta$, the density matrix is $|\mathrm{G}\rangle\langle\mathrm{G}| + |a\rangle\langle\mathrm{G}| + |\mathrm{G}\rangle\langle a|$, because the jump term maps $|a\rangle\langle\mathrm{G}|$ onto $\hat{c}|a\rangle\langle\mathrm{G}|\hat{c}^\dagger = 0$. The vector $|a\rangle$ therefore obeys

$$i\,\frac{d}{dt}|a\rangle \;=\; (\hat{H}_{\mathrm{eff}} - E_0)\,|a\rangle + \eta\cos(\omega_d t)\,\hat{X}|\mathrm{G}\rangle, \tag{9}$$

where $E_0$ is the ground-state energy, and its steady state is $|a\rangle = |A_+\rangle e^{-i\omega_d t} + |A_-\rangle e^{i\omega_d t}$ with

$$|A_\pm\rangle \;=\; \frac{\eta}{2}\,(\pm\omega_d - \hat{H}_{\mathrm{eff}} + E_0)^{-1}\,\hat{X}|\mathrm{G}\rangle. \tag{10}$$

The co-rotating output amplitude is $\langle\mathrm{G}|\hat{c}|A_+\rangle$. The amplitude of the photon displacement, $\langle\mathrm{G}|\hat{b}|A_+\rangle + \langle A_-|\hat{b}|\mathrm{G}\rangle$, gives the same peak separation, and a propagation of the full Lindblad equation at $\kappa = 0.02$ a.u. reproduces Eq. (10).

With an emitter linewidth $\gamma_e$, that is, $\omega_e \to \omega_e - i\gamma_e/2$, the transmission maxima of the two-level model move apart as $\kappa$ grows, by approximately $\kappa\gamma_e/8g^2$ relative. For $\gamma_e = 0.1g$, the separation increases by 1.2% at $\kappa = g$ and 3.5% at $\kappa = 3g$, while the pole splitting $2\sqrt{g^2 - (\kappa - \gamma_e)^2/16}$ decreases once $\kappa > \gamma_e$.

The molecular response is the co-rotating amplitude of the dipole operator along $\boldsymbol{\lambda}$ under the same drive, $\langle\mathrm{G}|\hat{d}|A_+\rangle + \langle A_-|\hat{d}|\mathrm{G}\rangle$. In the two-level model its maxima merge at $\kappa = 2\sqrt{2}\,g$, which

corresponds to $Q = 5.8$ for $H_2$. In the exact model, its peak separation decreases about twice as much as the pole splitting, and at $Q = 5$ it has a single maximum (Table S9). Both peak separations are converged in the photon-number cutoff $n_{\max} = 8$.

**Minimum gap over the cavity frequency**

Ref. 11 defines the Rabi splitting as the minimum gap over a scan of the cavity frequency. We repeated the exact calculations with this definition, scanning $\omega$ from 0.50 to 0.62 Ha on a grid of $10^{-4}$ Ha followed by bounded refinement. The lossless minimum gap lies slightly above the resonant frequency of the main text and is 0.08% smaller than the gap there. With the averaging time fixed at the value that gives $\kappa\tau = 0.688$ at the resonant frequency, the minimum gap of Eq. (7) decreases by 17.36%, compared with 17.32% at fixed frequency, and the minimum pole splitting decreases by 0.232% at $Q = 60$, compared with 0.228%.

**$H_2O$**

For $H_2O$/6-31G, the time-averaged Hamiltonian reduces the QED-CCSD-22 gap by 17.5% at $\kappa\tau = 0.688$ and by 25.3% at $\kappa\tau = 1$, almost exactly as for $H_2$, whereas the pole splitting from the scan of Sec. S7 decreases by 0.47% at $Q = 154$ and by 58% at $Q = 15$, close to the exceptional point.

**p-Nitroaniline**

The p-nitroaniline calculations use the geometry and the cc-pVDZ basis of Ref. 11, with all electrons correlated, and $\lambda = 0.05$ a.u. along the long molecular axis, which is the $C_2$ axis. Only the $A_1$ block of the Jacobian couples to the cavity, and its three lowest roots are computed by a Davidson solver in this block (9.4 of 69.8 million QED-CCSD-22 amplitudes), with products of the Jacobian and a vector from central differences of the residual (step $10^{-4}$) and a residual-norm threshold of $10^{-6}$. The complex stationary state is converged to $10^{-9}$ at each loss rate, starting from the lossless state. The lossless polaritons lie at 4.307 and 5.607 eV, and the next $A_1$ root at 6.99 eV.

At each quality factor, the decrease of the pole splitting is about three quarters of the resonant two-level value (Table S10).

Table S10: Complex polariton roots of p-nitroaniline (QED-CCSD-22/cc-pVDZ, $\lambda = 0.05$ a.u., $\omega = 5.660$ eV) versus the quality factor: relative change of the pole splitting, the resonant two-level value $(1 - \kappa^2/4\Omega_0^2)^{1/2} - 1$ with the lossless splitting $\Omega_0 = 1.300$ eV, and the linewidths $\gamma_k = -\mathrm{Im}\,\omega_k$.

| $Q$ | $\kappa$ (meV) | pole | two-level | $\gamma_{\mathrm{LP}}$ (meV) | $\gamma_{\mathrm{UP}}$ (meV) | $\gamma_k/(\kappa/2)$ |
|---|---|---|---|---|---|---|
| 60 | 94.3 | −0.049 % | −0.066% | 25.7 | 21.0 | 0.545, 0.445 |
| 20 | 283 | −0.45% | −0.59% | 77.2 | 63.0 | 0.545, 0.445 |
| 10 | 566 | −1.80% | −2.40% | 155 | 126 | 0.546, 0.445 |
| 5 | 1132 | −7.4% | −10.0% | 310 | 251 | 0.548, 0.443 |

For the time-averaged Hamiltonian of p-nitroaniline, we use QED-CCSD-21, the method of Ref. 11, and write the one-electron dipole self-energy as in the eT program, $1/2\,(\boldsymbol{\lambda}\cdot\widehat{\boldsymbol{d}})\,\boldsymbol{S}^{-1}(\boldsymbol{\lambda}\cdot\widehat{\boldsymbol{d}})$ with the per-electron dipole $\boldsymbol{\lambda}\cdot(\boldsymbol{r} - \sum_A Z_A\,\boldsymbol{R}_A/N)$. With the self-energy scaled as in Eq. (7), the resonance moves to lower cavity frequency, and near it, at 5.430 eV, the gap decreases by 11.06% with the cc-pVDZ basis (Table S11). The decrease depends little on the frequency, and the 6-31G basis gives nearly the same changes.

Table S11: Polariton gap of the time-averaged Hamiltonian of p-nitroaniline, Eq. (7) (QED-CCSD-21, $\lambda = 0.05$ a.u., $Q = 60$, $\tau = 5.01$ fs), relative to the lossless gap at 5.660 eV in the same basis, and the one-photon weights $w$ of the normalized right eigenvectors of the lower and upper polaritons, which are nearly equal near resonance. The last row is the cc-pVDZ value of Ref. 11.

| Hamiltonian | $\omega$ (eV) | gap (eV) | change | $w_{\mathrm{LP}}$, $w_{\mathrm{UP}}$ |
|---|---|---|---|---|
| cc-pVDZ | | | | |
| lossless | 5.660 | 1.250 | | 0.62, 0.50 |
| Eq. (7) | 5.430 | 1.111 | −11.06% | 0.60, 0.55 |
| 6-31G | | | | |
| lossless | 5.660 | 1.317 | | 0.59, 0.54 |
| Eq. (7) | 5.480 | 1.175 | −10.79% | 0.55, 0.61 |
| | 5.430 | 1.168 | −11.33% | 0.57, 0.59 |
| Ref. 11, lossless | 5.660 | 1.250 | | |